\documentclass[12pt]{article}

\usepackage{graphicx}
\graphicspath{{./figures/}}
\usepackage{dcolumn}
\usepackage{bm}
\usepackage{amsfonts}
\usepackage{amsthm}
\usepackage{amsmath}
\usepackage{amssymb}
\usepackage{slashed}
\usepackage{color}
\usepackage{textcomp}
\usepackage{subfig}

\def\be{\begin{equation}}
\def\ee{\end{equation}}
\def\ba{\begin{eqnarray}}
\def\ea{\end{eqnarray}}

\def\bdm{\begin{displaymath}}
\def\edm{\end{displaymath}}
\def\la{~\mbox{\raisebox{-.6ex}{$\stackrel{<}{\sim}$}}~}
\def\ga{~\mbox{\raisebox{-.6ex}{$\stackrel{>}{\sim}$}}~}
\def\bq{\begin{quote}}
\def\eq{\end{quote}}

 at 10truept

\newcommand{\bea}{\begin{eqnarray}}
\newcommand{\eea}{\end{eqnarray}}

\newcommand{\bi}{\begin{itemize}}
\newcommand{\ei}{\end{itemize}}

\newcommand{\beq}{\begin{equation}}
\newcommand{\eeq}{\end{equation}}
\newcommand{\beqa}{\begin{eqnarray}}
\newcommand{\eeqa}{\end{eqnarray}}

\def\la{~\mbox{\raisebox{-.6ex}{$\stackrel{<}{\sim}$}}~}
\def\ga{~\mbox{\raisebox{-.6ex}{$\stackrel{>}{\sim}$}}~}

\def\ltap{\ \raise.3ex\hbox{$<$\kern-.75em\lower1ex\hbox{$\sim$}}\ }
\def\gtap{\ \raise.3ex\hbox{$>$\kern-.75em\lower1ex\hbox{$\sim$}}\ }
\def\gl{\ \raise.5ex\hbox{$>$}\kern-.8em\lower.5ex\hbox{$<$}\ }
\def\roughly#1{\raise.3ex\hbox{$#1$\kern-.75em\lower1ex\hbox{$\sim$}}}

\begin{document}

\thispagestyle{empty}
\begin{flushright}
June 2018
\end{flushright}
\vspace*{1.5cm}
\begin{center}
{\Large \bf Irrational Monodromies of Vacuum Energy}
\vskip.3cm

\vspace*{1.26cm} {\large Nemanja Kaloper\footnote{\tt
kaloper@physics.ucdavis.edu}}\\
\vspace{.5cm} {\em Department of Physics, University of
California, Davis, CA 95616, USA}\\

\vspace{1.5cm} ABSTRACT
\end{center}
We present a theory with axion flux monodromies coupled to gravity, that reduces to 
the local vacuum energy sequester below the axion mass
scales. If the axion potentials include
a term generated by nonperturbative couplings to gauge sectors, with a decay constant incommensurate with monodromy periods, the low energy potential germinates a landscape of irrational axion vacua, with arbitrarily small cosmological constants. The sensitivity of the values of cosmological constants to unknown UV physics can be greatly reduced. The variation of the cosmological constant in each vacuum, from one order in perturbation theory to the next, can be
much smaller than the na\"ive cutoff. The nonperturbative transitions in the early universe between the vacua populate this landscape, similar to the case of irrational axion. In such a landscape of vacua a small cosmological constant
can naturally emerge.

\vfill \setcounter{page}{0} \setcounter{footnote}{0}
\newpage

\section{Introduction}

There are two sides to the cosmological constant problem \cite{zeldovich,wilczek,wein}. From the quantum field theory (QFT) point 
of view, the problem arises because the bare cosmological constant term must cancel the divergent vacuum energy  pieces \cite{wein, dreitlein,linde,veltman}. The divergences reappear at every order of perturbation theory, and the counterterm must change at each order by large amounts. 
The counterterm is not an observable: the renormalized cosmological constant is a sum of the vacuum energy loops and the counterterm, and so is a UV-sensitive quantity. 
Yet the instability of the counterterm in the loop expansion indicates 
the real problem: the observable is very sensitive to any unknown physics in the UV. If a QFT contains heavy fields just above a chosen cutoff, when they are included and then integrated out, the counterterm must change by a lot to keep the renormalized $\Lambda$ much smaller than the cutoff. Thus the problem of UV sensitivity: the cosmological 
constant, only sourcing gravity in the IR is sensitive to unknown UV physics whose other effects in the IR are
greatly suppressed.

An entirely different issue is why is the physical cosmological constant so 
small \cite{zeldovich,wein}, $\Lambda \la 10^{-12} \, {\rm eV}^4$. 
So miniscule a $\Lambda$ seems {very odd, in particular when we frame it in the context of {\it technical naturalness}: the vanishing of $\Lambda$ does not appear to lead to an emergence of any new symmetry.} The Poincare and de Sitter symmetry groups are different, but their ranks are the same. So from the viewpoint of a QFT on the background geometry, these can be viewed as deformations of one another, and there does not seem to be any symmetry enhancement when $\Lambda = 0$. Thus with the standard lore, the observed smallness of $\Lambda$ does not appear to be due to protective effects of a hidden symmetry\footnote{Softly broken conformal symmetry might help suppress some of the corrections, but it doesn't completely remove large terms and in addition it needs a very light dilaton \cite{dilatons}. On the other hand, the cancellation might occur thanks to supersymetry which might only be broken globally \cite{witten}.}. The problem appears particularly grave when the observations are compared with the theory, leading  to a discrepancy of some 60 to 120 orders of magnitude\footnote{Being unable to even estimate the error adds extra layers of embarassment.}. {This means that even if one knows the full UV completion of the theory, one does not have a mechanism in QFT which correlates the full vacuum
energy of the theory to the counterterm, that would guarantee that their sum is much smaller than the 
individual contributions.} 
Clearly, to address this problem one must go beyond QFT.

The full theory of quantum gravity, described by a path integral involving integration over geometries, could provide some insight into these puzzles. Attempts to tease these out of the path integral using simplified geometries and the saddle point approximation have been variously made \cite{Hawking:1984hk}, but these approximations may not be under control \cite{Klebanov:1988eh, Duff:1989ah}, and are deemed unreliable. Since the cosmological constant thus appears unconstrained from first principles, with QFT analyses suggesting that the very large values are favored but nothing precluding small values, one might in turn try to address to problem using  an {\it a posteriori} selection of $\Lambda$ based on the fact that the universe exists, is very old and large, and apparently conducive to the emergence of complex chemistry and, ultimately, life. Applications of the Anthropic Principle to address the cosmological constant problem, outlined in
\cite{Linde:1984ir,Weinberg:1987dv,Vilenkin:1994ua}, require the presence of a landscape: a framework which admits many low energy vacua, all supporting the same or very similar low energy QFTs but with a diversity of small cosmological constants, and a mechanism to populate this landscape
\cite{Banks:1984cw,Abbott:1984qf,Brown:1987dd,Banks:1991mb,Bousso:2000xa,Feng:2000if,Donoghue:2000fk}. With these mechanisms, the cosmological constant just might have to be selected by a dice toss, with the posterior condition that there is somebody to toss the dice.

Regardless of one's prejudice about anthropics, it is plausible that the observed value of $\Lambda$ is linked to the cosmological initial conditions. These do not appear to be very restrictive. In fact they may also be
uncertain: a relativistic cosmology governed by general relativity (GR) which evolves in time performs a scanning in energy. It starts at highest energies accessible to QFT and evolves towards ever smaller energies. 
The terminal values of $\Lambda$ 
set in the IR as a boundary condition will  extrapolate back to some UV region with a given cutoff. If the cutoff changes, this region may need to change as well. Because each region dominated by $\Lambda$ grows extremely large even the smallest  differences in $\Lambda$ at the initial time will result in exponential corrections to the volume later on\footnote{This argument is not very covariant, but we believe that it illustrates the issue at least at scales where semi-classical gravity governs the evolution of an isotropic and homogeneous universe at large scales.}. This can add difficulty to defining probabilistic measures in cosmology, starting with the colloquial volume measure, and may therefore affect likelihood estimates for other observables. 

On the other hand, recently the author and collaborators have pursued the approach of vacuum energy sequester,
which involves an addition of Lagrange multipliers to the gravitational sector of the theory\footnote{Similar ideas have been noted previously \cite{andreimult,tseytlin,HT}. Other applications of $4$-forms with topological sectors are given in \cite{Dvali:2001sm}.} \cite{KPglobal,KPlocal,KPbubbles,KPgrav,etude}. 
The idea is to promote the `couplings' in the gravitational sector, which are UV-sensitive quantities (i.e., the cosmological constant, the Planck scale, $\ldots$) into Lagrange multiplier fields. These fields are forced to be constant by introducing their couplings to {\it non-gravitating} $4$-form field strengths\footnote{Such as $\int\Lambda dA$, where $A$ is a $3$-form potential. The gauge symmetry of $A$ ensures that $d\Lambda = 0$.}. The field equations for the Lagrange multipliers relate them to the standard fields, selecting 
solutions which automatically absorb arbitrary loop corrections in
perturbation theory. Specifically, the Lagrange multiplier $\Lambda$ is the cosmological constant counterterm just as in \cite{HT}, and it cancels the QFT vacuum energy loops automatically as a solution of the equations of motion. The non-gravitating $4$-forms $F = dA$ pick the loops up \cite{KPlocal,KPgrav}, but do not
backreact on the geometry since their stress energy tensor is identically zero \cite{nicolai}. {These terms 
behave as an independent, {\it second measure} for the effective QFT coupled to (semi)classical gravity, which differ
from $\sqrt{g}$ off-shell, which modifies the gravitational field equations in the infinite wavelength limit.
The modification is designed to remove the total vacuum energy, classical+quantum+counterterm, from 
the total stress energy tensor sourcing local gravity. This violates the Equivalence Principle, but only in 
the limit which is not accessible to observations\footnote{It may be that the observed smallness of the cosmological 
constant is the only directly observable effect.}. As a result the mechanism provides a workaround for the 
Weinberg no-go theorem \cite{wein} which blocks the cancellation of the quantum corrections to
vacuum energy in the conventional frameworks of QFT coupled to gravity, in the absence of supersymmetry and/or conformal symmetry.} 

While such dynamics
does remove the QFT vacuum energy loops from the stress-energy tensor \cite{KPlocal,KPgrav}, similarly to the ideas suggested in \cite{selft,degrav}, it leaves a residual cosmological constant which is completely arbitrary. {The residual cosmological constant can
take any value below the QFT cutoff; however,  once a value is assigned any unknown UV physics corrections never affect it.} 
This obviously raises the question about what may be used to determine the residual
$\Lambda$, and select its value to be small {to fit the observations.} 

In the absence of a full, UV complete framework, an obvious and conservative (sic!) direction is to build a landscape of possible vacua, and explore how it gets populated. The purpose of this communication is to show how to taxonomize the end results of vacuum energy sequester, by embedding the sequester mechanism in a theory with two axion flux monodromies \cite{KS,NDA} coupled to gravity. {The reason for utilizing monodromies is twofold. First, flux monodromies  in the dual magnetic formulation 
automatically realize the modified measure terms utilized by sequester dynamics.
These terms arise due to the auxiliary terms required to ensure that the $4$-form fluxes are constant, reflecting
the fact that $4$-forms in $4D$ cannot propagate on their own. Even when the local dynamics is included by the
addition of a scalar field which mixes with the $4$-form, yielding its mass term, the residual gauge symmetries --- 
specifically, the discrete shift symmetry --- preserve the structure of the Lagrangian required for sequestering vacuum energy. Secondly, the monodromy structures may 
provide hints for how to extend the sequester mechanism beyond QFT coupled to (semi)classical gravity, which we still
pursue here for simplicity's sake\footnote{Providing an example of a QFT where the cosmological constant
is UV-insensitive and small, even with classical gravity, would nevertheless be a unique example at the present
time, even if we ignore the graviton loops. As Weinberg's no-go \cite{wein} shows, such vacuum selections cannot
occur in standard EFT coupled to gravity.}. The monodromies are realized by two scalar fields which mix 
bilinearly with the $4$-forms, and are `eaten' by them in the axial gauge, providing them with a local
degree of freedom \cite{KS}.}  One of the monodromies has the axion shift symmetry broken at a 
very high scale, which fixes the axion vev at a value close to the Planck scale. 
The other can be broken at a lower scale, allowing the second axion to be lighter. If this axion also 
couples to a gauge theory whose nonperturbative effects induce an additional axion potential, with a decay constant that is an irrational multiple of monodromy periods, the low energy potential at energies below the light axion mass spans a landscape of irrational axion vacua \cite{Banks:1991mb}. These vacua have vacuum energies which can be arbitrarily small. Their values can be much less 
sensitive to unknown UV physics, because they are controlled by the scale of the nonperturbative gauge theory potential which could be much smaller than the cutoff. 

On the other hand, like in the irrational axion \cite{Banks:1991mb}, the differences between the vacuum energies can also be arbitrarily small. This follows because the ratios of periods of potentials are irrational numbers, and so the overall potential is not periodic, instead approaching any allowed value arbitrarily closely. Nonperturbative transitions between the vacua, which can be described simply by both axion tunneling and by the emission of membranes charged under the $3$-form fields, populate this landscape. Since the energy differences can be arbitrarily small thanks to incommensurability of the periods, there will be many vacua\footnote{Similar observations in a different context were made in \cite{andreifernando}.} with very small vacuum energy like in \cite{Banks:1991mb}. Crucially, unlike \cite{Bousso:2000xa}, this {\it does not} require a large number of fields and forms. 

If the nonperturbative corrections to the potential which induce this fine structure of vacua is sufficiently suppressed by instanton effects, it could even happen that all the vacua have a small cosmological constant in the range of the observed one. If the nonperturbative effects are less suppressed, the cosmological constant at the observed level can be accommodated by invoking the Anthropic Principle. {Which of these options is realized clearly depends
on the details of UV completion.} 

Our mechanism above seems to critically rely upon the concept of the 
irrational axion  \cite{Banks:1991mb}  in order to populate the phase space of the theory with many
vacua where the residual cosmological constant is small. If the axion really had couplings with irrational 
ratios to different sectors in fundamental theory -- at least in simple realizations of the scenario -- the model
would have an emergent global shift symmetry. This seems to be in conflict with ideas about the role 
of symmetries in quantum gravity, which are presumed to be either gauged or broken \cite{Banks:2010zn}. 
If this were the case, one may legitimately wonder if our scenario is merely an example of an unrealistic toy model,
without a chance for a UV completion. 

However, the various low energy vacua do not need to have cosmological constants which differ from each other 
by arbitrarily small amounts. It suffices to have differences which are {\it small enough} -- i.e. 
$\Delta \Lambda \la 10^{-12} ~ {\rm eV}^4$. If so, then the low energy effective potential for the axion 
which partakes in sequester only needs to approximate the irrational potential for a finite range of
field variation. A related option is that the ratio of the charges controlling the axion potential is a 
rational number $p/q$, 
but $p$ and $q$ are very large mutually prime numbers. A scenario like this was recently considered in
\cite{Bachlechner:2015gwa}. In this case, the low energy values of the cosmological constant develop a band spectrum as in crystals, with the fine structure of the levels controlled by $p/q$, that can be dense enough
to approximate the irrational axion.

In either case, we can imagine that the global symmetry of the axion sector, resembling the irrational axion,
is {\it emergent} a low energies. In truth, it is a remnant of a gauge symmetry which was broken at some
high scale  (but below the scale of quantum gravity) to a residual discrete gauge symmetry. Axion monodromies provide a clue
of how this may happen, and at this point are not ruled out by any fundamental reason. At the effective 
theory level, then, such potentials can be obtained by coupling the axion to a 4-form sector with a nonzero flux
in some compact dimension, and a fundamental anomalous gauge theory, and then dimensionally reducing.
The compactification breaks gauge symmetry; after stabilizing the volume moduli, the low energy effective potential for the axion will approximate
the irrational axion for a range of values of $\phi$. This is because the couplings to different fundamental sectors
have different dependence on volume moduli terms. In effect these are the {\it finite} renormalizations
of the gauge couplings at ow energies which can be different because they receive different IR 
corrections even if their UV limits coincide. A similar assumption 
is made in the string landscape paper \cite{Bousso:2000xa} where the membranes charged under different 
4-forms have charges which are mutually irrational in 4D. When the axion varies out of this window of values,
the theory would decompactify due to the backreaction  on the stabilizing sources, and the emergent irrationality would disappear. A more precise analysis of such scenarios would be very interesting. We intend to return to this 
question elsewhere.

The paper is organized as follows. In section \ref{sec2} we will review the local theory of vacuum energy sequester in terms of new variables, which make the connection with axion monodromies more straightforward. In section \ref{sec3} we consider the extension of local vacuum energy sequester to include the monodromies and discuss the ensuing landscape of low energy vacua. We outline the cosmology and the anthropic selection of $\Lambda$ in section \ref{sec4}. We consider the scanning processes and the early universe cosmology, when the landscape is populated. Finally we summarize in section \ref{sec5}. We speculate about using this landscape for anthropically addressing the smallness of $\theta_{QCD}$ in the appendix. 

\section{Local Vacuum Energy Sequester}
\label{sec2}

The local vacuum energy sequester action of \cite{KPlocal} is given by
\be
S = \int d^4 x \sqrt{g} \left[ \frac{\kappa^2(x)}{2} R  - \Lambda(x) - {\cal L}_m( g^{\mu\nu} , \Phi) + \frac{1}{4!}\frac{\epsilon^{\mu\nu\lambda\sigma}}{\sqrt{g}} \left(\sigma(\frac{\Lambda(x)}{\mu^4}) F_{\mu\nu\lambda\sigma} +\hat \sigma(\frac{ \kappa^2(x)}{ M_{Pl}^2}) \hat F_{\mu\nu\lambda\sigma} 
\right) \right] 
 \, . ~~~~~~~~
\label{actionJloc}
\ee
Here, $F_{\mu\nu\lambda\sigma} = 4 \partial_{[\mu} A_{\nu\lambda\sigma]}$ and $\hat F_{\mu\nu\lambda\sigma} = 4 \partial_{[\mu} \hat A_{\nu\lambda\sigma]}$ are the two 4-forms whose gauge symmetries render the Lagrange multipliers $\Lambda$ and $\kappa^2$ constant on shell, respectively. {In principle, we could have used any
other pseudoscalar which is independent of $\sqrt{g}$ off shell in place of $\epsilon F, \epsilon \hat F$. Note that the metrics in the terms 
which depend on $F, \hat F$ completely cancels out. We resort
to $4$-forms because they provide the simplest means of introducing another measure, and because they 
naturally arise from monodromy models \cite{KS}, that might point towards UV completions.} The functions 
$\sigma$ and $\hat \sigma$ are differentiable functions of dimensions $({\rm mass})^2$, but otherwise arbitrary. The normalization scales $\mu \la M_{Pl}$ are the QFT and gravity cutoffs. {They are required by dimensional reasons, and are determined by the 
UV completions of the QFT and gravity sectors, respectively, such that $\Lambda \le \mu^4$ and
$\kappa^2 \le M^2_{Pl}$. If the theory is to be fully technically natural, the cutoffs should also be subject to
$\mu^2 \le M_{Pl}^2/N$, where $N$ is the number of QFT degrees of freedom in the IR.} 
The $\epsilon$ in the last two terms is the standard Levi-Civita symbol. The matter sector $\Phi$ is described by the Lagrangian ${\cal L}_m(g^{\mu\nu}, \Phi)$. Due to the metric cancellations in the last two terms, they are purely topological, and their stress tensor vanishes. 

To make contact with monodromies easier, let us field-redefine the variables in (\ref{actionJloc}). We introduce 
new scalar fields $m \phi = \sigma(\Lambda/\mu^4)$ and $M \hat \phi = \hat \sigma(\kappa^2/M^2_{Pl})$. Since
$\sigma$s are differentiable, we can invert these functions and write $\Lambda = \mu^4 V(\frac{m \phi}{\mu^2})$, 
$\kappa^2 = M_{Pl}^2 {\cal U}(\frac{M \hat \phi}{M_{Pl}^2})$, where $V, {\cal U}$ are dimensionless functions with Taylor expansions. The masses $m$ and $M$ are arbitrary parameters here, {which are
obtained from folding together the cutoffs $\mu, M_{Pl}$ with in-principle arbitrary dimensionless normalizations 
of the numerical coefficients in the expansion of $\sigma, \hat \sigma$. Specifically, the masses can be smaller
than the cutoffs when the functions $\sigma,\hat \sigma$ are steep \cite{KS}. Since $\sigma, \hat \sigma$ are radiatively stable \cite{KPglobal,KPlocal}, this choice is not UV sensitive.} In terms of the new variables  the action becomes\footnote{Because of the coupling $\sim \hat \phi \epsilon \hat F$, $\hat \phi$ must be a pseudoscalar to preserve CPT. So we treat the coefficient of $R$ as a function of a pseudoscalar. However the theory (\ref{actmon1}) can be modified by replacing ${\cal U} \rightarrow {\cal U}(M\hat \chi/M_{Pl}^2)$ and $M \hat \phi \rightarrow \hat \phi \hat \chi$ where $\hat \chi$ is a scalar.}
\be
S = \int d^4 x \sqrt{g} \left[ \frac{M_{Pl}^2}{2} {\cal U}(\frac{M \hat \phi}{M_{Pl}^2}) R  - \mu^4 V(\frac{m\phi}{\mu^2})  - {\cal L}_m( g^{\mu\nu} , \Phi) +  \frac{\epsilon^{\mu\nu\lambda\sigma}}{4!\sqrt{g}} \left( m \phi F_{\mu\nu\lambda\sigma} +M \hat \phi \hat F_{\mu\nu\lambda\sigma} 
\right) \right] 
 \, . ~~~~~
\label{actmon1}
\ee
Clearly the first two terms resemble a (two)scalar-tensor gravity, while the last two terms are the scalar-$4$-form couplings enforcing the constancy of the scalars. { In other words, the extra gravitational scalar fluctuations are
projected out from the theory by virtue of the constraints induced by the $4$-form field strengths.} 

To see how the vacuum energy sequester works, let's look at the field equations from (\ref{actmon1}):
\ba
&&\partial_\mu \phi = \partial_\mu \hat \phi = 0 \, , ~~~~~~~  F_{\mu\nu\lambda\sigma} = \mu^2 V' \sqrt{g} \epsilon_{\mu\nu\lambda\sigma} \, , ~~~~~~~
  \hat F_{\mu\nu\lambda\sigma} =-\frac{{\cal U}'}{2} R \sqrt{g} \epsilon_{\mu\nu\lambda\sigma} \, ,  \nonumber \\
&& ~~~~~~~~~~~~~~~~~~~~~~~~~~~~~~~ M_{Pl}^2  {\cal U} \, G^\mu{}_\nu = T^\mu{}_\nu-\mu^4 V \delta^\mu{}_\nu  \, .\label{eoms}
\ea
Here  $T_{\mu\nu}=\frac{2}{\sqrt{g}}\frac{\delta }{\delta g^{\mu\nu}}\int d^4 x \sqrt{g}{\cal L}_m( g^{\mu\nu} , \Phi)  $ is the matter stress-energy tensor. Primes denote derivatives with respect to dimensionless arguments of the potentials. Since the Einstein's equation involves the function $V$ of an a priori arbitrary constant variable $\phi$, we should determine it by using the other equations. To isolate the constant scalar contributions in them, we 
first trace over them. This yields $M_{Pl}^2 {\cal U} R = 4\mu^4 V - T$. Then 
taking the worldvolume average of $R$ yields $\langle R \rangle = \int d^4 x \sqrt{g} R/\int d^4x \sqrt{g} = (4 \mu^4 V - \langle T \rangle)/M_{Pl}^2 {\cal U}$. Finally we eliminate $\langle R \rangle$ using the $4$-form equations, which 
give $\langle R \rangle = - 2 \frac{\mu^2 V'}{{\cal U}'} \int \hat F/\int F$. We find
\be
\mu^4 V = \frac{\langle T \rangle}{4} - \frac{ \mu^2}{2} \, M_{Pl}^2 {\cal U} \, \frac{V'}{{\cal U}' } \, \frac{\int \hat F}{\int F}   \, , ~~~~~ M_{Pl}^2 {\cal U} G^\mu{}_\nu = T^\mu{}_\nu- \frac{\langle T \rangle}{4} \, \delta^\mu{}_\nu +\frac{ \mu^2}{2} \, M_{Pl}^2 {\cal U} \, \frac{V'}{{\cal U}' } \, \frac{\int \hat F}{\int F}  \, \delta^\mu{}_\nu \, .
\label{Veq}
\ee
{ We stress here that the term $\propto \langle T \rangle$ is very different than any superficially similar terms 
which arise in local modifications of gravity. First and foremost, this term is a {\it constant} with a fixed overall normalization, instead of the local term $T = T^\mu{}_\mu(x^\nu)$, which could appear in e.g. scalar-tensor 
gravity. This is denoted by the brackets $\langle Q \rangle = \int d^4x \sqrt{g} Q(x)/\int d^4x \sqrt{g}$. This term is {\it nonlocal}, but it only appears as an on-shell solution of the totally local field equations (\ref{eoms}). Further
this term does not involve any arbitrary integration constant, such as appears in the unimodular formulation 
of general relativity. Here, the counterterm of the theory is completely fixed at any order in
perturbation theory.} It cannot be picked randomly, but must be selected to solve the equations of motion. 

Note that as long as the vacuum energy contains only internal matter lines, the term $\langle T \rangle/4$ exactly cancels the vacuum energy contributions from $T^\mu{}_\nu$ at {\it any loop order} due to the general covariance of the theory \cite{KPglobal,KPlocal}, and replaces it with the radiatively stable term $- \frac{ \mu^2}{2} \, M_{Pl}^2 {\cal U} \, \frac{V'}{{\cal U}' } \, \frac{\int \hat F}{\int F} $. 
Indeed, consider the matter sector configurations which correspond to the minima of the matter action,
$\partial_\Phi V_{matter} =0$. In this case, $T^\mu{}_\nu = - V_{matter}(min) \, \delta^\mu{}_\nu$ and so
\be
\mu^4 V = -V_{matter}(min) - \frac{ \mu^2}{2} \, M_{Pl}^2 {\cal U} \, \frac{V'}{{\cal U}' } \, \frac{\int \hat F}{\int F}   \, , ~~~~~ M_{Pl}^2 {\cal U} \, G^\mu{}_\nu =  \frac{ \mu^2}{2} \, M_{Pl}^2 {\cal U} \, \frac{V'}{{\cal U}' } \, \frac{\int \hat F}{\int F}  \, \delta^\mu{}_\nu \, .
\label{Veqmin}
\ee
The counterterm replaced the QFT vacuum energy $V_{matter}(min)$ with a term $\propto  V' = m \partial_\phi V$! This will automatically repeat at any order in the loop expansion \cite{KPlocal}. { The only remaining constant term sourcing gravity in the vacuum will be the term $\propto \int \hat F/\int F$, but this term is completely 
separate from the matter sector cosmological constant,. Thus this means that the matter sector cosmological
constant, ie the sum classical+quantum+counterterm, does not gravitate. It is completely sequestered away. 
Basically this happens because of the chain of equations of motion 
 $ - 2 \frac{\mu^2 V'}{{\cal U}'} \int \hat F/\int F = \langle R \rangle = (4 \mu^4 V - \langle T \rangle)/M_{Pl}^2 {\cal U}$, connecting on shell radiatively stable quantity $ - 2 \frac{\mu^2 V'}{{\cal U}'} \int \hat F/\int F $ with $(4 \mu^4 V - \langle T \rangle)/M_{Pl}^2 {\cal U}$. This
violates the equivalence principle for constant sources, ie in the limit of infinite wavelengths, but that does not
conflict with any observational bounds or observed phenomena.} 

The radiative stability of the remaining term follows after accounting for the renormalization of the Planck scale
$M_{Pl}^2 {\cal U}$,
\be
\left(M^{ren}_{Pl}{}\right)^2 \simeq M_{Pl}^2 {\cal U}+ {\cal O}(N) 
\left({M}_{UV}^{}\right)^2 + \sum_{species} {\cal O}(1)  
m_{M}^2 \ln({M}_{UV}^{}/m_{M}) + \sum_{species} {\cal O}(1)  
m_{M}^2 + ...\, ,
\ee
where ${M}_{UV}^{} \simeq \mu$ is the matter UV regulator, $N$ counts the matter degrees of freedom,
and $m_{M}$ a mass of a virtual particle in the loop\footnote{This coupling will be natural when $M_{Pl}^2 \sim N \mu^2$.}. Once the quadratically divergent $M_{Pl}^2 {\cal U}$ is set to its experimentally determined value given by the Planck scale -- which corresponds to the choice of $\hat \phi$ -- it is radiatively stable as long as 
$\mu \la M_{Pl}/\sqrt{N}$, as we noted above, and required in \cite{KPlocal}. The same holds for ${\cal U}'$, by naturalness of ${\cal U}$.
The integrals in $\int \hat F/\int F$ are fixed by the classical boundary conditions for the $4$-forms, and are dominated by IR physics thanks to the huge domains of integration. 
While this ignores graviton multiplet loops\footnote{See \cite{KPgrav} for a modification that may cancel the graviton loops in addition to matter loops.} it nevertheless accomplishes to stabilize the matter sector cosmological constant in the limit where gravity is used to detect it. 

Why is this happening? How can the counterterm function $\mu^4 V$ know what value to pick to cancel the QFT vacuum energy contributions automatically? The trick lies in the fact that gauge invariant variables included in the covariant actions involve the world-volume integral $\int \sqrt{g}$. In standard GR, this global degree of freedom completely factorizes from the action, and is left to be determined purely by the boundary conditions. Thus the cosmological constant counterterm, which is the Legendre dual of this variable, remains completely arbitrary. While the renormalization procedure introduces the UV boundary condition for it, in order to cancel the divergence, what remains is completely undetermined. In addition, if the UV boundary condition is changed, for example by moving the cutoff around, the counterterm needs to be changed {\it by hand}. This is the usual radiative instability. 

In the sequester theory (\ref{actmon1}) this variable does not factor out due to the topological terms. Instead, 
$\int\sqrt{g}$ remains present in the $4$-form equations, and is extracted out by their integration.  Removing it from the field equations requires the division of the $4$-form fluxes which in turn fixes the counterterm 
$\mu^4 V$ uniquely. The remainder, 
\be
\Lambda_{residual} = 
- \frac{ \mu^2}{2} \, M_{Pl}^2 {\cal U} \, \frac{V'}{{\cal U}' } \, \frac{\int \hat F}{\int F} \, ,
\label{remainder}
\ee
is a product of a classical term, given by the ratio of fluxes, the radiatively stable, renormalized effective Planck
scale $M_{Pl}^2 {\cal U}$ and the ratio of the derivatives $V'/{\cal U}'$. As long as $V'$ is a flat function, this term
will remain stable under loop corrections, since ${\cal U} \sim {\cal O}(1)$, and in the absence of tunings also ${\cal U}' \sim {\cal O}(1)$. In principle, $\Lambda_{residual}$ could even be zero for some values of $\langle T \rangle$,
but it suffices that once it is chosen flat, it remains flat over a wide range of $\phi$. Since its value is completely arbitrary, and as long as the overall scale of $\Lambda_{residual}$ is suppressed relative to
$M_{Pl}^4$, any further reduction must come from tuning $V'$ and/or the ratio of fluxes. Yet, once set, this tuning will not be destabilized by quantum corrections.

{ Note that in principle we could have integrated out the `constants' $\phi, \hat \phi, F, \hat F$ from the start.
The resulting effective action would be a very  nonlocal modification of gravity, which however would not introduce any new local degrees of freedom, but only modify the way how vacuum energy gravitates. This is clear from our formulation based on Lagrange multipliers, and the fact that integrating those out cannot change the canonical structure of the theory. On the other hand, if such an action were to be truncated to only few leading terms, the spurionic local degrees of freedom
may appear, contaminating the resulting truncation with pathologies. We therefore choose to work with the
Lagrange multipliers, which manifestly separate the dynamics at finite wavelengths from the dynamics in the infinite wavelength limit --- ie, local fluctuations from the cosmological constant. 

We then see that this mechanism evades Weinberg's no-go because it involves different local measures, using  the topological 4-form terms. These do not generate sources of stress energy, and so violate strongly the equivalence principle in the infinite wavelength limit, replacing $V$ with $V'$ in (\ref{remainder}). The violation of the equivalence principle in this way is not a problem: such cases abound, with any axion-gauge field coupling being 
an example: $\phi \, Tr(G \, \, ^*G)$ is topological, without a stress energy tensor.
As the $4$-forms soak up the QFT vacuum energy loops, they screen them from gravity at infinite wavelengths, and fix the
residual cosmological constant to (\ref{remainder}), which however is completely arbitrary so far.} 

\section{Monodromies and Sequester}
\label{sec3}

The topological terms in the action (\ref{actmon1}) are some of the terms which arise in the flux monodromy 
constructions of large field inflation, coupling massless axions and non-dynamical  $4$-forms \cite{KS,NDA}. It has been argued \cite{KS,NDA} that such models are low energy effective theories of generic monodromy models such as  \cite{eva} and many others. It has also been argued that monodromies which stem from topological couplings are very generic in string compactifications \cite{uranga}. This raises a \ question: could the vacuum energy sequester be embedded in a field theory with a monodromy (monodromies)? The answer is affirmative. Pursuing it we will find the dynamics behind the ``stationary cancellation" of QFT vacuum energy enforced by the sequester constraints. The structure which arises from linking sequester with flux monodromies is a landscape of vacua which are scanned by the fluxes and axion vevs.

{ First, let us give here a lightning review of the main features of flux monodromy \cite{KS,NDA} which are pertinent for vacuum energy sequester. Starting with a massless $4$-form, we can readily rewrite it in terms of its magnetic dual, ${\cal Q} = \frac{1}{4! \sqrt{g}} \epsilon^{\mu\nu\lambda\sigma} F_{\mu\nu\lambda\sigma}$, by adding the  
constraint ${\cal Q} (F_{\mu\nu\lambda\sigma} - 4 \partial_{[\mu} A_{\nu\lambda\sigma]})$ directly in the action and integrating $F$ out: 
\be
- \int d^4 x \sqrt{g} \frac{1}{48} F^2_{\mu\nu\lambda\sigma} ~~~ \rightarrow ~~~ - \int d^4x \sqrt{g} \frac12 {\cal Q}^2 
+ \frac{1}{24} \int d^4x {\cal Q} \epsilon^{\mu\nu\lambda\sigma} {\cal F}_{\mu\nu\lambda\sigma} \, ,
\label{monoact}
\ee
where ${\cal F}_{\mu\nu\lambda\sigma} = 4\partial_{[\mu} A_{\nu\lambda\sigma]}$ is the residual $4$-form
enforcing $d{\cal Q} = 0$. This is precisely the structure of the local constraints employed in vacuum energy sequester. Now, in general, since the theory contains membranes charged under $A$, the flux variable $Q$ 
is quantized, and can only change from region to region by an integer multiple of the unit charge. The variation
can be made continuous, however, by the introduction of a massless Goldstone mode which mixes with the
original $4$-form. This is done by adding $- \frac12 (\partial \phi)^2$ and $\frac{m}{24} \phi \epsilon^{\mu\nu\lambda\sigma} F_{\mu\nu\lambda\sigma}$ to the left hand side of Eq. (\ref{monoact}), which 
yields
\ba
&& \int d^4 x \sqrt{g} \Bigl(- \frac{1}{48} F^2_{\mu\nu\lambda\sigma}  - \frac12 (\partial \phi)^2 + \frac{m}{24} \phi \epsilon^{\mu\nu\lambda\sigma} F_{\mu\nu\lambda\sigma}  \Bigr) ~~~ \rightarrow ~~~~~~~~~~~~~~~~~~~~ \nonumber \\
&& ~~~~~~~~~~~~~~~~~~  \int d^4x \sqrt{g} \Bigl( - \frac12 (\partial \phi)^2 - \frac12 (m\phi+ {\cal Q})^2 
+ \frac{1}{24} \int d^4x {\cal Q} \epsilon^{\mu\nu\lambda\sigma} {\cal F}_{\mu\nu\lambda\sigma} \, .
\label{monoact1}
\ea
Note that the effective potential for $\phi$ depends only on the linear combination $m\phi + {\cal Q}$, as required
by the discrete gauge shift symmetry ${\cal Q} \rightarrow {\cal Q} + q$, $\phi \rightarrow \phi - {\cal F}$ where
$ {\cal F} = q/m$ \cite{KS,NDA}. Moreover, if additional irrelevant operators are included in the theory, such as eg
$\propto F^{2n+2}/M^{4n}_{UV}$, where $M_{UV}$ is the UV cutoff, modifying $-\frac{1}{48} F^2$ to some 
general $K(F^2)$, the gauge symmetries ensure that after dualization the effective potential is 
$V(m\phi + {\cal Q})$, where $V$ is the Legendre transform of $K$ \cite{Dvali:2001sm,KS,NDA}. 
In other words, the discrete shift symmetry remains unbroken --- as it must, being a gauge symmetry. 
Thus
the potential $V$ is very general. An important constraint however is that near the minima the 
effective mass squared must be positive, which corresponds to the requirement that the dual $4$-form
must have the kinetic term of the usual sign as opposed the opposite sign. In the dual $4$-form language
this follows from the fact that if the membrane nucleation rate is not completely suppressed, the 
wrong sign $4$-form quadratic term yields a vacuum instability in the very late universe \cite{sandora}. 
}

Next, let us add two such flux monodromy structures to the sequester theory (\ref{actmon1}), and `deconstruct' the `counterterm potential' $\mu^4 V$. We start with 
\ba
S &=&\int \sqrt{g} \, \Bigg[ \frac{M_{Pl}^2}{2} {\cal U}(\frac{M \hat \phi}{M_{Pl}^2}) R  - {\cal L}_m( g^{\mu\nu} , \Phi) + \frac{1}{4!} \frac{\epsilon^{\mu\nu\lambda\sigma}}{\sqrt{g}} \left( m \phi F_{\mu\nu\lambda\sigma} +M \hat \phi \hat F_{\mu\nu\lambda\sigma} \right) ~~~~~~~~~~~~~~~~  \nonumber \\
&&~~~~~ - \frac12 (\partial \phi)^2 - \frac12 (\partial \hat \phi)^2 - \frac{1}{48} F_{\mu\nu\lambda\sigma}^2  
- \frac{1}{48} \hat F_{\mu\nu\lambda\sigma}^2 + (\frac{\phi}{ f} + \theta) \, \frac{g^2}{16\pi^2} \, Tr( G\, \, ^*G )+ ... \Bigg]\,. 
\label{fluxmonos}
\ea
The first line of (\ref{fluxmonos}) is precisely the local sequester (\ref{actmon1}) sans the potential $\mu^4 V$, whereas the second line contains the terms which complete the topological terms into mondoromies\footnote{Even if we hadn't added the kinetic term for $\hat \phi$, it would have gotten it from mixing with $R$.}, and also a coupling of the axion $\phi$ to a gauge theory, represented by the $\propto \phi \, Tr(G \, \, ^*G)$ term. 
Here $\theta$ is an arbitrary phase, which arises from some fermion mass matrix and as in \cite{Banks:1991mb} we assume that there are no global symmetries in the fermion sector which can be used to remove it. 

While we are not explicitly writing the couplings of the form  fields to membranes charged under them in (\ref{fluxmonos}), we do assume that they are present, with elementary charges $q, \hat q$. This leads to the compactification of axion fields $\phi, \hat \phi$ and can be seen easily by replacing the $4$-forms with their magnetic duals, which come in as Lagrange multipliers in terms like 
$\propto {\cal Q} \epsilon^{\mu\nu\lambda\sigma} (F_{\mu\nu\lambda\sigma} - 4 \partial_{[\mu} A_{\nu\lambda\sigma]})$ \cite{KS,NDA}. This step simply repeats the dualization from our lightning review. 
Adding one each for $F,\hat F$ to (\ref{fluxmonos}), completing the squares and integrating $F$s out yields
\ba
S &=&  \int \Bigg\{ \sqrt{g} \, \Bigg[ \frac{M_{Pl}^2}{2} {\cal U}(\frac{M \hat \phi}{M_{Pl}^2}) R  - {\cal L}_m
- \frac12 (\partial \phi)^2 - \frac12 (\partial \hat \phi)^2 - \frac{m^2}{2} (\phi + \frac{\cal Q}{m})^2  
- \frac{M^2}{2} (\hat \phi + \frac{\hat {\cal Q}}{M})^2 \Bigg] ~~~~
\nonumber \\
&&~~~~~  + \frac16 \epsilon^{\mu\nu\lambda\sigma} \Big({\cal Q} \partial_{[\mu} A_{\nu\lambda\sigma]} +\hat{\cal Q} \partial_{[\mu} \hat A_{\nu\lambda\sigma]}  \Big) + (\frac{\phi}{ f} + \theta) \, \frac{g^2}{32\pi^2} \, \epsilon^{\mu\nu\lambda\sigma} \, Tr( G_{\mu\nu}\, \, G_{\lambda\sigma} )+ ... \Bigg]\,. 
\label{fluxdualmons}
\ea
Here we have explicitly separated metric from topological terms. Now, { since the $4$-form fluxes are
quantized in the units of membrane charge, ${\cal Q} = Nq$, $\hat{\cal Q} = \hat N \hat q$, where $N, \hat N$ count the number of membranes sourcing the fluxes, we see that shifting $\phi \rightarrow \phi - {\cal F}$ and ${\cal Q} \rightarrow {\cal Q} + q$ is a discrete gauge symmetry of (\ref{fluxdualmons}) as long as the $\phi, \hat \phi$ periods satisfy ${\cal F} = q/m$ and $\hat {\cal F} = \hat q/M$.} This gives rise to the monodromy structures\footnote{The gravitational prefactor ${\cal U}(M\hat \phi/M_{Pl}^2)$ 
breaks the shift symmetry of $\hat \phi$ and significantly alters the structure of the monodromy vacua for $\hat \phi$. However, that is acceptable since a heavy field $\hat \phi$ never moves much, as we will discuss shortly.}
for $\phi, \hat \phi$. Again, we repeat that even if $F^2$ terms were replaced by more general kinetic terms
$K(F^2), \hat K(\hat F^2)$, similar duality transformations can still be made, replacing the quadratic effective potentials for dual variables by 
\be
\frac{m^2}{2} (\phi + \frac{\cal Q}{m})^2  ~~~ \to ~~~ V(\frac{m \phi + {\cal Q}}{{\cal M}^2/4\pi}) \, ,
~~~~~~~~~~~ \frac{M^2}{2} (\hat \phi + \frac{\hat {\cal Q}}{M})^2 ~~~ \to ~~~ \hat V(\frac{M \hat \phi + \hat {\cal Q}}{\hat {\cal M}^2/4\pi}) \, , 
\ee
where ${\cal M}, \hat {\cal M}$ are the cutoffs controlling the corrections to the $4$-form 
theories \cite{Dvali:2001sm,KS,NDA}, and $V, \hat V$ are Legendre transforms of $K, \hat K$. The key difference between the quadratic potentials and more general ones is that the monodromy energy branches as a function of $\phi$ are more complicated than parabolas, 
with minima where $V, \hat V$ can have any nonnegative value\footnote{Negative values may be problematic because of stability, since they would imply branches with negative kinetic terms for the $3$-forms $A, \hat A$, as noted in \cite{sandora}.}. They would retain the same periods ${\cal F}, \hat {\cal F}$. Note that these potentials, once computed, are
radiatively stable \cite{KS,NDA} as long as their arguments are at most of order unity, $\frac{m \phi + {\cal Q}}{{\cal M}^2/4\pi} \la {\cal O}(1)$, $\frac{M \hat \phi + \hat {\cal Q}}{\hat {\cal M}^2/4\pi} \la {\cal O}(1)$. The normalizations
here are a consequence of Na\"ive Dimensional Analysis as discussed in \cite{NDA}.

The action (\ref{fluxmonos}) suggests that despite the absence of fermion sector global symmetries, the parameter 
$\theta$ can nevertheless be absorbed away by shifting $\phi \rightarrow \phi -f \theta$, since naively the only
term that would change under the shift is $m \phi \, \epsilon F \rightarrow m \phi \, \epsilon F -  mf \theta \, \epsilon F$, and the variation $mf \theta \epsilon F$ is a total derivative. However, shifting $\phi$ changes the energy of the system because the monodromies spontaneously break shift symmetry. Due to periodicity of $\phi$, continuous shift symmetry is broken down to a discrete one, and $\theta$ can only be removed exactly if 
$l {\cal F}/f  + \theta = 2\pi n$ for some integers $l,n$. If $\frac{{\cal F}}{2\pi f} = \sigma$ is rational this will only be possible for special values of $\theta$. On the other hand, if $\frac{{\cal F}}{2\pi f}$ is irrational, we can always pick integers 
$l,n$ such that $\theta +2\pi (l \sigma - n)$ is arbitrarily close to zero \cite{Banks:1991mb,niven}. We will take 
this case here and work with irrational $\sigma$ (working in what follows with $\omega = 2\pi \sigma$). This means that the states with different $\theta$ in (\ref{fluxdualmons}) are perturbatively degenerate. 

{ This degeneracy will be lifted when the axion is coupled to an additional nonabelian gauge theory at scales where the gauge theory strong coupling effects kick in. These will yield nonperturbatively induced corrections
to the axion effective potential, just like in QCD, which will induce a fine structure to the effective potential of the theory. Indeed, we imagine that the axion $\phi$ is coupled to some hidden sector nonabelian gauge theory 
which is strongly coupled at a high scale $\bar \lambda$. The corrections to the $\phi$ effective action at 
scales below $\bar \lambda$, induced by the nonperturbative gauge dynamics, can be estimated using the dilute 
instanton gas approximation, and will break spontaneously the axion discrete gauge shift symmetry. At energies below the scale $\bar \lambda$ of the gauge theory $G$ we can integrate out the gauge fields due to confinement, along with any other heavy degrees of freedom. At the leading order in the dilute instanton gas approximation\footnote{The trigonometric form of the potential is not essential; the point is that the potential is periodic and differentiable. We will work with the trigonometric form for convenience's sake.} the induced  nonperturbative correction to the potential is $\tilde V = \lambda^4 [1-\cos(\omega \phi/{\cal F} + \theta)] $\,. Note that $\lambda$ is not the same as the symmetry breaking scale $\bar \lambda$. It could be much lower, depending on the gauge group topology and the details of the fermion sector charged under it, which can yield exponential suppressions, 
\be
\lambda \simeq \bar \lambda \, e^{-S/4} \, ,
\label{stronginst}
\ee
where $S$ is the instanton action, $S \gg 1$. This does not occur in the standard QCD, where
$\lambda^4 \simeq m^2_\pi f^2_\pi \simeq \Lambda_{QCD}$, since the pion mass and decay constant
are close to the QCD scale. However with a different fermion contents, one can find examples where 
$\lambda \simeq  \lambda \, e^{-S/4} \ll \bar \lambda$, such as in quintessence constructions \cite{milliev}.}
The total effective potential for $\phi$ below the scale $\bar \lambda$ is
\be
V_{tot} =  V(\frac{m \phi + {\cal Q}}{{\cal M}^2/4\pi})  + \lambda^4\Big[1-\cos(\frac{\omega \phi}{\cal F} + \theta)\Big] + \ldots \, .
\label{effpot}
\ee
This potential has global minima where it almost vanishes when $\omega$ is irrational  \cite{Banks:1991mb,vafawitten}. Note that the mass of $\phi$ is $\sim m^2$ above the scale $\bar \lambda$, while below it the axion is heavier, with mass $\sim m^2 + \omega^2 \lambda^4/{\cal F}^2 > m^2$. Moreover note that this term in general spontaneously breaks the discrete shift symmetry of $\phi$. It is crucial that this breaking is very soft (in our case that is guaranteed by using nonperturbative effects to induce the breaking) in order to maintain the flatness
of the monodromy potential. Otherwise, dangerous operators will be induced in the loop expansion\footnote{For example, if this discrete shift symmetry were broken arbitrarily, we could not forbid operators like $\phi^2 F^2/M^2 \sim F^4/m^2 M^2$ with ${\cal O}(1)$ prefactors, which are disastrous for the flatness of the $\phi$ potential \cite{KS}.}. 

We will ignore such structures in the $\hat \phi$ sector, restricting for simplicity's sake to the quadratic $\hat V = M^2 (\hat \phi + \hat {\cal Q}/M)^2/2$. Perhaps the most unusual term in (\ref{fluxmonos}) is \, ${\cal U}(M \hat \phi/M_{Pl}^2)$. From the point of view of low energy physics, this term must be  $\sim 1$ in order to reproduce reasonable phenomenology and cosmology. From our discussion of sequester above, this is also its natural value, that renormalization will drive it to. The effective potential for the canonically normalized field $\chi \sim M_{Pl} \ln \sqrt{\cal U}$ in the Einstein frame is
\be
\hat V_{eff} = \frac{\hat V}{{\cal U}^2} + \ldots \, .
\label{eframepot}
\ee
The mass of this field is given by
\be
m_\chi^2 \sim \partial_\chi^2 V_{eff} \sim \frac{M_{Pl}^2}{M^2} \frac{\cal U}{{\cal U}'} \partial_{\hat \phi} \Bigl[\frac{\cal U}{{\cal U}'} \partial_{\hat \phi}  \Bigl( \frac{\hat V}{{\cal U}^2} \Bigr)  \Bigr] \sim M_{Pl}^2\, .
\label{chimass}
\ee
This means that the field $\hat \phi$ is for all practical intents and purposes decoupled from the action, with its
fluctuations integrated out. Yet its background value will be slightly deformed by the presence of
gravitational sources in the theory, which move the field slightly away from its minimum, displacing it by 
$\Delta \hat \phi \propto R$. In other words, the potential ${\cal U}$ forces $\hat \phi$ to behave just like a Lagrange
multiplier. { This is absolutely {\it crucial} for the vacuum energy sequester: the background value of $\hat \phi$
does respond ever so slightly when the source for gravity changes, yielding an equation of motion which forces
a constraint on the zero  mode of $R$. This is of course the second set of constraint equations from the local
vacuum energy sequester, that is essential for fixing the cosmological constant counterterm to precisely the value
which cancels the quantum vacuum energy. Essentially, the heavy field $\hat \phi$ plays a role very similar to the 
flattening heavy fields in string monodromy constructions \cite{exercises}, which yield small changes of both the
location and the form of the light field potential minima.

Let us now show in detail how all this works in the theory based on (\ref{fluxdualmons})\footnote{Some of the technicalia on sequestering monodromies were identified in discussions with A. Padilla.}. } We are interested in energy scales below the Planck scale. Since $\hat \phi$, as we have seen, is fixed by Planck scale effects, we can integrate it out and only keep its background. As we noted it gets slightly deformed by sources. In light of this it is convenient to rewrite (\ref{fluxdualmons}) as
\ba
S &=&  \int \Bigg\{ \sqrt{g} \, \Bigg[ \frac{M_{Pl}^2}{2} {\cal U}(\frac{M \hat \phi}{M_{Pl}^2}) R  - {\cal L}_m
- \frac12 (\partial \phi)^2 - V_{tot}(\phi, Q) 
- \frac12 M^2 (\hat \phi + \frac{\hat {\cal Q}}{M})^2 \Bigg]
\nonumber \\
&&~~~~~~~~~  + \frac{1}{4!} \epsilon^{\mu\nu\lambda\sigma} \Big({\cal Q} F_{\mu\nu\lambda\sigma} +\hat{\cal Q}  \hat F_{\mu\nu\lambda\sigma}  \Big) + ... \Bigg\}\, ,
\label{fluxdualmoneff}
\ea
where we reintroduced\footnote{At first sight this may seem redundant; after all these $3$-forms are auxiliary in
flux monodromy constructions. However, a closer look shows -- as we will see explicitly in what follows -- that
these structures do not simply decouple since they mix with matter sources. The situation is analogous to what happens with sources in theories with kinetic and mass mixings of $U(1)$ fields, where in the propagation eigenbasis the currents carry charges of both $U(1)$'s \cite{holdom}.} topological $4$-forms $F_{\mu\nu\lambda\sigma} = 4 \partial_{[\mu} A_{\nu\lambda\sigma]}$ and $\hat F_{\mu\nu\lambda\sigma} = 4 \partial_{[\mu} \hat A_{\nu\lambda\sigma]}$. 
Here the potential $V_{tot}$ below the scale $\bar \lambda$ is given by (\ref{effpot}). Note that (\ref{fluxdualmoneff}) resembles (\ref{actmon1}), although it contains three Lagrange multipliers $\hat \phi, {\cal Q}, \hat {\cal Q}$ and a dynamical axion $\phi$. The field equations for scalars and forms are 
\ba
&&\partial_\mu {\cal Q} = \partial_\mu \hat {\cal Q}  = 0 \, , ~~~~~~~ \nabla^2 \phi = - \partial_\phi V_{tot} \, , ~~~~~~~~~~ 2 M(\hat \phi + \frac{\cal Q}{M}) = \, {\cal U}' R  \, , \nonumber \\
&& ~~~~~ F_{\mu\nu\lambda\sigma} =  \partial_{\cal Q} V_{tot} \sqrt{g} \epsilon_{\mu\nu\lambda\sigma} \, , ~~~~~~~~\,
  \hat F_{\mu\nu\lambda\sigma} =  M\Bigl({\hat \phi} + \frac{\hat {\cal Q}}{M} \Bigr) \sqrt{g} \epsilon_{\mu\nu\lambda\sigma} \, , 
  \label{eomsnew}
\ea
whereas the gravitational equations are
\ba
M_{Pl}^2 {\cal U} \, G^\mu{}_\nu &=& M_{Pl}^2 \Bigl(\nabla^\mu \nabla_\nu - \delta^\mu{}_\nu \nabla^2 \Bigr) \, {\cal U} - \partial^\mu \phi \partial_\nu \phi -  \frac12 (\partial \phi)^2 \delta^\mu{}_\nu  \nonumber \\
&& + T^\mu{}_\nu - \Bigg( V_{tot} +  \frac12 M^2 (\hat \phi + \frac{\hat {\cal Q}}{M})^2 \Bigg)\delta^\mu{}_\nu  + \ldots 
\label{eomsnewgr}
\ea
We have kept the scalar derivatives in (\ref{eomsnewgr}) for completeness. However they play no role in the cancellation of QFT vacuum energy corrections and the selection of the residual cosmological constant, as we will now explain. 

Let us first consider the derivatives of  $\hat \phi$ and ${\cal U}$. The $\hat \phi$ dependence of ${\cal U}$ can be disentangled from gravity by a conformal transformation which maps (\ref{eomsnew}) to the Einstein conformal frame, with the canonically normalized scalar $\chi \sim \ln \sqrt{\cal U}$. As we showed in 
(\ref{chimass}), the fluctuations of $\chi$ (or $\hat \phi$) are too heavy to matter, with Planckian mass, and can be
safely integrated out. In fact we could completey integrate $\hat \phi$ out, replacing it by the solution of its equation of motion since it is so heavy (\ref{chimass}). This would replace ${\cal U}(M\hat\phi/M^2_{Pl})$ by a series in the powers of $R$ and $\hat {\cal Q}$. The resulting theory would produce exactly the same dynamics as when we treat the background vev $\hat \phi$ as a Lagrange multiplier, and so we will ignore the details here. 

So ignoring the fluctuations of $\hat \phi$, the background $\hat \phi$ is found by minimizing 
the effective potential (\ref{eframepot}), found by tracing (\ref{eomsnewgr}) and substituting $R$ in the third of Eq. (\ref{eomsnew}), { which yields $R = \frac{2M}{{\cal U}'} (\hat \phi + \hat {\cal Q}/M)$. Ignoring spacetime derivatives, and using the trace equation of the Einstein's equations, $R = 2 \frac{(M \hat \phi + \hat {\cal Q})^2}{M_{Pl}^2 {\cal U}} + \frac{4V_{tot} - T}{M_{Pl}^2 {\cal U}}$, we find by combining the two,
\be
\frac{(M \hat \phi + \hat {\cal Q})}{{\cal U}^2} - \frac12 \, \frac{{\cal U}'}{{\cal U}^2} \Bigg(2 \frac{{(M\hat \phi + {\cal Q})^2}}{M_{Pl}^2 \, {\cal U}} + \frac{4V_{tot} - T}{M_{Pl}^2 \, {\cal U}} \Bigg)  =  \partial_\chi V_{eff} =  0 \, .
\label{backscalareq}
\ee
This equation is precisely the statement that the canonically normalized scalar field 
$\chi \sim M_{Pl} \ln \sqrt{\cal U}$ is residing in its minimum because it is so heavy. Notice that if the 
minimum coincides with a minimum of ${\cal U}$, then at the minimum $M \hat \phi + {\cal Q} = 0$. Alternatively,
and more generally, taking the coefficients in the Taylor expansion of ${\cal U} \,$ to be ${\cal O}(1)$ and also taking\footnote{The challenge is keeping scalars light, making them heavy is easy.} $M \simeq M_{Pl}$, all we need is a root $\hat \phi_0$ of 
(\ref{backscalareq}) with all parameters ${\cal O}(1)$ in the units of Planck scale, and with $ V_{eff}/{\cal U}^2$ being 
convex at that root. Note that the choice of such a root automatically determines $\hat F$. This is {\it not} a fine tuning. It is merely a statement that $\hat F$ is a {\it solution} of the field equations (\ref{eomsnew}), (\ref{eomsnewgr}), and once $\hat \phi_0$ is determined for a fixed $\hat {\cal Q}$, $\hat F$ is uniquely fixed. 
Further note that due to
the periodicity of the monodromy structure, there may be many such roots. 

Having selected a root, which is to serve the role of the vacuum expectation value for $\hat \phi$, when $R$ deviates from the background value above, we can determine 
how much it distorts $\hat \phi_0$ by expanding the third equation of (\ref{eomsnew}) around $\hat \phi_0$. This yields 
\be
\frac{M\delta \hat \phi}{M_{Pl}^2} = \frac12 \, {\cal U}'(\hat \phi_0) \, \frac{\delta R}{M_{Pl}^2} \,  \Bigl[1 + {\cal O}
(\frac{\delta R}{M_{Pl}^2})\Bigr] \, , 
\label{phivar} 
\ee 
making it clear that the distortion is always small and the corrections are tiny when $R < M_{Pl}^2$. 
Since we are steering clear from Planckian scales (where we'd have to deal with full-blown quantum gravity, not only large semiclassical corrections) we can drop all the spacetime derivatives of $\hat \phi$ -- i.e. ${\cal U}$ -- from the right hand side of (\ref{eomsnewgr}), since they are of order $\nabla^2 \delta R/M_{Pl}^2$, and therefore suppressed relative to the Einstein terms by two powers of the curvature in the units of Planck scale. Yet this small distortion of $\hat \phi$ is very important for the $4$-form sector since it feeds in the flux of $\hat F$, as we see when we combine the last equations in the two lines of (\ref{eomsnew}). We will return to this shortly. Note, that
the discussion so far serves only to fix the {\it vev} of $\hat \phi$. We haven't yet gotten to the point where
we can say what the value of the cosmological constant is for that $\hat \phi_0$.   }

Now we turn to $\phi$. In (\ref{fluxdualmons}), $\phi$ behaves pretty much like an ordinary matter field,
except that it couples directly to the monodromy field ${\cal Q}$. This gives it a mass protected by the 
discrete gauge symmetry $\phi\rightarrow \phi - {\cal F}$, ${\cal Q} \rightarrow {\cal Q} + q/m$, ${\cal F}m = q$. Hence in perturbation theory the loops involving $\phi$ could generate operators that depend on $\partial \phi$ and 
$\phi +{\cal Q}/m$. These modify the effective potential away from the simple quadratic form, but preserve
the dependence on the arguments \cite{KS,NDA}. The fact that $\phi$ is massive helps with understanding the dynamics of sequester. 

{ To proceed we need to evaluate the constant Lagrange multipliers ${\cal Q}, \hat {\cal Q}$. To this end we trace the gravitational equations and
evaluate their (regulated) spacetime averages.} With the equations (\ref{eomsnew}), (\ref{eomsnewgr}) it is clear that this will include the averages of the derivative terms like
$$
\langle \nabla^2 \phi \rangle = \frac{\int \sqrt{g} \, \nabla^2 \phi}{\int \sqrt{g}} \, , ~~~~~~~~~~~~~~~
\langle (\partial \phi)^2 \rangle = \frac{\int \sqrt{g} \, (\partial \phi)^2 }{\int \sqrt{g}} \, ,
$$
and possibly similar expressions involving higher order corrections. 
The integrals are over the whole cosmic history of the universe\footnote{These integrals in reality are merely the `observational tools' required to measure the cosmological constant to sufficient precision. In principle, taking a region of the spacetime to be large but finite, e.g. $\sim 100$ times the size of the current horizon should be enough. We will ignore the distinction here.}.
Since $\phi$ is massive it will settle in the minimum of its potential, and its derivatives
will asymptotically vanish. So in the expressions for the spacetime averages of derivatives,
the numerators will receive a contribution for only a subset of the domain of integration. Similarly when averaging time-evolving potentials, we can replace them with their asymptotic values. Again, the variations will occurs over finite regions, and be bounded. As the IR regulators of the spacetime integrals are removed, the averages vanish. Hence we will use only the vacuum field configurations and Einstein's equations for averaging, which isolates 
QFT vacuum energy corrections and the cosmological constant counterterm. The relevant simplified equations are
\ba
&& 2 M(\hat \phi + \frac{\hat{\cal Q}}{M}) = \, {\cal U}' R  \, , ~~~ F_{\mu\nu\lambda\sigma} =  \partial_{\cal Q} V_{tot} \sqrt{g} \epsilon_{\mu\nu\lambda\sigma} \, , ~~~
  \hat F_{\mu\nu\lambda\sigma} =  \frac{{\cal U}'}{2} R \sqrt{g} \epsilon_{\mu\nu\lambda\sigma} \, ,  \nonumber \\ 
&& ~~~~\, \partial_\phi  V_{tot} = 0 \, ,  ~~~~~~~\,~~ M_{Pl}^2 {\cal U} \, G^\mu{}_\nu = T^\mu{}_\nu - \Bigg( V_{tot} +  \frac12 M^2 (\hat \phi + \frac{\hat {\cal Q}}{M})^2 \Bigg)\delta^\mu{}_\nu   \, , 
\label{eomsnewgrsim}
\ea
where ${\cal Q}, \hat {\cal Q}$ are constants. We have used the first equation to rewrite the third. Now,
averaging the trace $R$ yields 
\be
2 M^2 \langle (\hat \phi + \frac{\hat {\cal Q}}{M})^2 \rangle/M^2_{Pl} {\cal U} \, + (4\langle V_{tot} \rangle - \langle T\rangle)/M^2_{Pl} {\cal U} -  \langle R \rangle = 0 \, . 
\label{avr}
\ee

{ Next by using the first equation 
on the right hand side of (\ref{eomsnewgrsim}) we obtain $ \frac12 M^2 (\hat \phi + \frac{\hat {\cal Q}}{M})^2 = ({\cal U}' )^2 \, R^2/8$, which means that the terms $ (\hat \phi + \frac{\hat {\cal Q}}{M})^2$ and $\langle (\hat \phi + \frac{\hat {\cal Q}}{M})^2 \rangle$ are equalized rapidly throughout the cosmic history once the heavy field $\hat \phi \simeq {\rm const.}$ settles in the vacuum, and can be safely cancelled against each other, $ (\hat \phi + \frac{\hat {\cal Q}}{M})^2 - \langle (\hat \phi + \frac{\hat {\cal Q}}{M})^2 \rangle = 0$. Eg, $M \sim M_{Pl}$ is above the scale of inflation, so 
the field $\hat \phi$ will rapidly fall in its minimum, and its oscillations around $\hat \phi_0$ will be exponentially 
suppressed within a few efolds. So substituting $ \langle (\hat \phi + \frac{\hat {\cal Q}}{M})^2 \rangle$ in place of $ (\hat \phi + \frac{\hat {\cal Q}}{M})^2$ in the last of Eq. (\ref{eomsnewgrsim}), tracing the equation and subtracting a quarter of Eq. (\ref{avr}) from the diagonal part of the
gravitational equations we find 
\be
M_{Pl}^2 {\cal U} \, G^\mu{}_\nu = T^\mu{}_\nu - \frac14 \langle T \rangle \delta^\mu{}_\nu - \Bigg( V_{tot} - \langle V_{tot} \rangle \Bigg)\delta^\mu{}_\nu  - \delta^\mu{}_\nu \frac14 {M_{Pl}^2 {\cal U} } \langle R \rangle + \ldots \, , 
\ee
where ellipses denote higher order corrections, but also evolutionary contributions from excitations above the vacuum. These do not spoil radiative stability and are automatically small in old large universes \cite{KPglobal,KPlocal}. We eliminate $\langle R \rangle$ from this equation by 
integrating and dividing $4$-forms and using the third of (\ref{eomsnewgrsim}). This gives
$\langle R \rangle = \frac{2 \partial_{\cal Q} V_{tot}}{{\cal U}'} {\int \hat F}/{\int F}$. Substituting,
\be
M_{Pl}^2 {\cal U} \, G^\mu{}_\nu = T^\mu{}_\nu - \frac14 \langle T \rangle \delta^\mu{}_\nu - \Bigg( V_{tot} - \langle V_{tot} \rangle \Bigg)\delta^\mu{}_\nu  - \delta^\mu{}_\nu \frac{M_{Pl}^2}{2} \frac{\cal U}{{\cal U}'} \, \partial_{\cal Q} V_{tot} \, \frac{\int \hat F }{\int F }
+ \ldots \, .  
\label{eqs}
\ee
The equations (\ref{eqs}) are remarkably similar to the gravitational equations in (\ref{Veq}). As there, the term 
$\propto \langle T \rangle$ cancels QFT vacuum energy energy contributions, and $\langle V_{tot} \rangle$ cancels $ V_{tot} $ whenever $\phi + {\cal Q}/m$ settles in the minimum in which QFT vacuum energy corrections are calculated. That leaves the residual cosmological constant
\be
\Lambda_{residual} = \frac{M_{Pl}^2}{2} \frac{\cal U}{{\cal U}'} \, \partial_{\cal Q} V_{tot} \, \frac{\int \hat F }{\int F } \, , 
\label{rescc}
\ee
which is formally radiatively stable as before, as long as $V_{tot}$ is a flat function and once the Planck scale is renormalized and assigned its observed value. Note that as the UV corrections are absorbed order by order,
by changing $m \phi + {\cal Q}$, $\partial_{\cal Q} V_{tot}$ does change as well. Note also, that in principle 
the radiatively stable $\Lambda_{residual}$ could be large if $\partial_{\cal Q} V_{tot}$ is large.
}

A curious feature emerges when one considers the possible forms of the potential $V_{tot}$ which we mentioned above. Suppose for a moment that the gauge theory couplings in Eq. (\ref{effpot}) vanish. So $V_{tot}$ is only 
the monodromy term from (\ref{effpot}), $V_{tot} = V(\frac{m\phi + {\cal Q}}{{\cal M}^2/4\pi})$. Then $\partial_{\cal Q} V_{tot} = \partial_\phi V_{tot}/m$. But the fourth equation of (\ref{eomsnewgrsim}) asserts that $\partial_\phi V_{tot} = 0$ in any vacuum. Suddenly it seems that the residual cosmological constant of (\ref{rescc}) vanishes for any
vacuum of the theory!? 

A closer inspection of (\ref{eomsnewgrsim}) shows the problem with this: $\int F = \partial_\phi V_{tot} \int \sqrt{g}/m$ vanishes as well and so (\ref{rescc}) is indeterminate as it stands. But replacing $F$ by its on-shell form yields
\be
\Lambda_{residual} = \frac{M_{Pl}^2}{2} \frac{\cal U}{{\cal U}'} \, \langle \hat F \rangle \, .
\label{resccreg}
\ee
Since ${\cal U} \sim {\cal U}' \sim 1$, this means that $\Lambda_{residual} \sim M_{Pl}^2 \langle \hat F \rangle$, where $\langle F \rangle$ is an arbitrary spacetime average value of a classical  $4$-form flux. 
Instead of fixing the value of $\Lambda_{residual}$ to zero, the extension of vacuum energy sequester by the inclusion of monodromies alone made it a continuum. The actual observable cosmological constant can  now be anything, in principle.

This would be very similar to Brown-Teitelboim theory \cite{Brown:1987dd} in the limit of vanishing membrane charge. In that case, classical flux of the $4$-form, which screens the cosmological constant, can be discharged continuously, yielding all kinds of potential problems. 

Rather than revisiting this interesting case, we will side-step it here, and instead go to the case where $\phi$ couples to other sectors, and its potential receives additional contributions. This will change cosmic history and produce a different spectrum of residual cosmological constants. 

If we allow the coupling of $\phi$ to a gauge theory as in (\ref{fluxmonos}), and in turn get the low energy potential for $\phi$ like (\ref{effpot}), the expression for the residual cosmological constant (\ref{rescc})
is well behaved. Indeed, $\partial_{\cal Q} V_{tot}  = \partial_{\cal Q} V$, and $\partial_\phi V_{tot}  = m \partial_{\cal Q} V + \frac{\omega \lambda^4}{\cal F} \sin(\frac{\omega \phi}{\cal F} + \theta)$. Thus $\partial_\phi V_{tot} = 0$ implies 
\be
- \partial_{\cal Q} V = \frac{\omega \lambda^4}{m\cal F} \sin(\frac{\omega \phi}{\cal F} + \theta) \, .
\label{condition}
\ee
The residual cosmological constant becomes
\be
\Lambda_{residual} = - \frac{\omega}{2} \frac{\cal U}{{\cal U}'} \, \frac{M_{Pl}^2 }{m {\cal F} } \, \lambda^4 \, \sin(\frac{\omega \phi}{\cal F} + \theta)  \, \frac{\int \hat F }{\int F } \, . \label{resccregfine}
\ee
The dimensional normalization $\sim  \frac{M_{Pl}^2 }{m {\cal F} } \, \lambda^4$ is an energy scale 
set by the nonperturbative gauge theory dynamics, as we noted above. It can be much smaller than the cutoff, 
and it could even be very small \cite{milliev}. The dimensionless numbers are ${\cal U} \sim \, {\cal U}' \sim \int \hat F/\int F \sim {\cal O}(1)$ as we noted above. 

Since the field value $\phi$ changes by $\Delta \phi \la {\cal M}^2/4\pi m \gg M_{Pl}$ the argument of the sine in (\ref{resccregfine}) varies a lot as loop corrections and UV contributions are added:
$\omega \Delta \phi/{\cal F} \la \omega {\cal M}^2/4\pi {\cal F} m \gg 1$. Thus the sine oscillates wildly from one
step in the loop expansion to another. As a result, the residual cosmological constant $\Lambda_{residual}$ is
radiatively stable down to $\frac{M_{Pl}^2}{m {\cal F}} \lambda^4$ level, but unstabe below it. In other words introducing the discrete fine structure to the spectrum of the residual cosmological constant has reintroduced 
radiative instability, but at the scale that is much below the cutoff. This is very interesting, but by itself it might
not be enough. For a generic $\theta$ in (\ref{resccregfine}) this means that a typical
value of $\Lambda_{residual}$ is in fact $\frac{M_{Pl}^2}{m {\cal F}} \lambda^4$, that could be too large.

If $\theta = 0$ and $\omega \in \mathbf{N}$, implying that some fermions which are charged under the gauge group $G$ have remained massless, there would be values of $\phi$ where $\omega \phi/{\cal F} \equiv 2n \pi$,
and where the sine would vanish identically. However these may be widely spaced, and hence atypical, requiring fine tunings.

\begin{figure*}[thb]
\centering
\includegraphics[scale=0.41]{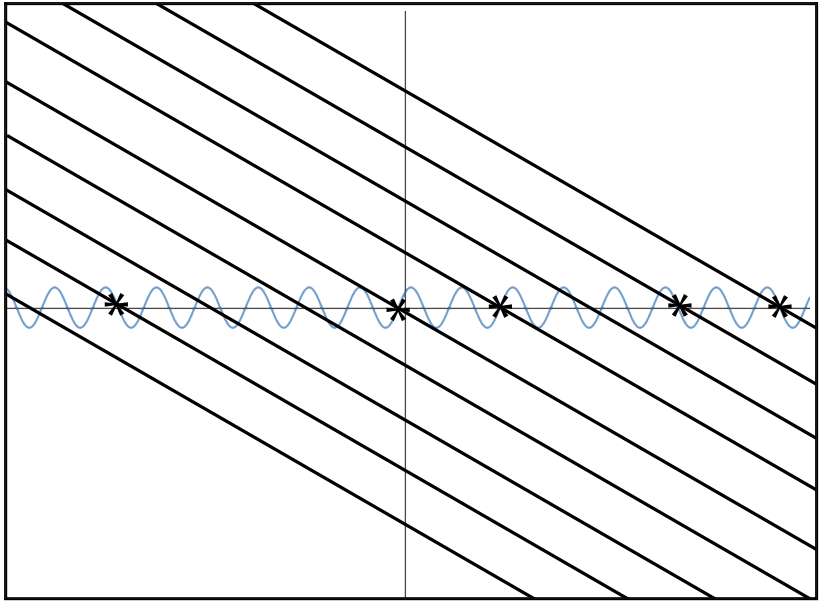}
\caption{Vacua are given by the intersections of the lines and the sine (\ref{condition}). For simplicity we took 
$V$ to be the quadratic monodromy here. The asterisks denote the vacua with very small residual cosmological constant due to $\omega \phi/{\cal F} + \theta \approx 2\pi n$. Alternatively,
if $\frac{M_{Pl}^2 }{m {\cal F} } \, \lambda^4 \la 10^{-12} \, {\rm eV}^4$, the cosmological constant is small in any intersection. }
\label{fig1}
\end{figure*}
Clearly, we would be amiss not to mention that the gauge theory to which $\phi$ couples might have the right topology and charged fermions with the right couplings, such that for a symmetry breaking scale $\bar \lambda$, 
due to the nonperturbative gauge theory dynamics, it so happens that  \cite{milliev}
\be
\frac{M_{Pl}^2}{m {\cal F}} \, \lambda^4 \simeq \frac{M_{Pl}^2}{m {\cal F}} \, \bar \lambda^4 e^{-S} \la 
10^{-12} \, {\rm eV}^4 \, ,
\ee
{ We stress that this small value of magnitude of the cosmological constant is radiatively stable due to its nonperturbative origin, and while it does not seem natural, is manufactured by nonperturbative gauge dynamics
without egregious fine tunings.} If so, both the value of $\Lambda_{residual}$ in (\ref{resccregfine}) and its radiative (in)sensitivity would drop to below the observed level of cosmological constant for any choice of $\phi$ and ${\cal Q}$. In such a case, the problem would be solved without any further work. In the cases of other gauge groups more work is needed to  produce such a small cosmological constant.

For this reason, we are taking $\omega$ to be irrational here, just in case. So the sine function has the period which is incommensurate with the monodromy period. Now, it is still true that for many of the vacua parameterized by $\phi$, the phase of the sine will not be arbitrarily close\footnote{In the extrema $\omega \phi/{\cal F} + \theta 
\approx \pi (2n+1)$, $\partial_\phi^2  V_{tot} \sim m^2 -
\frac{\omega^2 \lambda^4}{{\cal F}^2}$ and $\phi$ may be a tachyon. Those'd be maxima.}  to $2\pi n$ for any $n$. But due to the irrationality of $\omega$ every so often we will find that $\omega \phi/{\cal F} + \theta $ is arbitrarily close to $2\pi n$ for some $n$ \cite{niven}, as we explained previously. For those values of $\phi$ the residual cosmological constant will be tiny. This is illustrated in Fig. (\ref{fig1}). The resulting picture is the structure of vacua with a range of residual cosmological constants, whose magnitude can be pretty much any number $\la \frac{M_{Pl}^2 }{m {\cal F} } \, \lambda^4$.

Obviously, this is a landscape. The vacua are distributed randomly, with the overall scale of the cosmological constant set by $\frac{M_{Pl}^2 }{m {\cal F} } \, \lambda^4$. However, the QFT vacuum energy energy contributions 
larger than $\frac{M_{Pl}^2 }{m {\cal F} } \, \lambda^4$
are completely sequestered away from gravity, and the cosmological constant values are UV insensitive to 
physics at scales above  $\frac{M_{Pl}^2 }{m {\cal F} } \, \lambda^4$. Again, if this scale is $\la 10^{-12} \, {\rm eV}^4$, we could pick any vacuum and satisfy the observational bounds (with some more selection required to obtain
$\Lambda_{residual} \simeq 10^{-12} \, {\rm eV}^4$). When this scale is larger we need something else, and anthropics is the obvious recourse.

\section{Irrational Landscape of $\Lambda$} 
\label{sec4}

The irrational landscape of vacuum energy sequester which we have outlined above shares some of the features of the irrational axion landscape
\cite{Banks:1991mb} and the landscape permeated with large $4$-form fluxes \cite{Bousso:2000xa}. Like
\cite{Banks:1991mb} we also utilize the incommensurability of the periods of the $4$-form and gauge theory
contributions to the axion effective potential $V_{tot}(\phi)$, which generates a dense set of minima 
with arbitrarily small cosmological constant in many of them. At a technical level, \cite{Banks:1991mb} obtain
the potential for $\phi$ by coupling it to two different gauge theories whose periods are incommensurate.
In our case, we only need one gauge theory, generating the other contribution to the $\phi$ potential by
flux monodromy. Further, the differences between different values 
of the cosmological constant between different vacua can also be very small like in \cite{Banks:1991mb}. The main
difference between our scenario and the irrational axion is that the large quantum contributions and contributions from other sectors are completely sequestered away here, being diverted into the non-gravitating $4$-forms. The only contributions come from $\sim \frac{M_{Pl}^2 }{m {\cal F} } \, \lambda^4$ terms. 

In other words, we have outlined a landscape involving only renormalized, UV insensitive values
of vacuum energy in each vacuum, unlike the irrational axion \cite{Banks:1991mb}, where the higher order corrections and unknown UV physics can influence low energy predictions significantly. 
The net cosmological constant does scan densely the region of values near zero \underbar{regardless} 
of the regularized value $|\Lambda_{vacuum}|$.

As a result the subsequent landscape dynamics in our case is semiclassical, once the universe evolves 
past the end of eternal inflation. Like \cite{Bousso:2000xa} we can use eternal inflation to populate this landscape.  Let us take the discrete gauge shift symmetry breaking scale $\bar \lambda$ coming from the gauge theory to which
the axion couples 
to be above the scale of slow roll inflation\footnote{In contrast to \cite{Banks:1991mb} where the authors take the gauge theory to be QCD, and so $\lambda$ is much below the scale of inflation.}. In this case during inflation the instanton potential (the cosine) is already present, and it contributes to the cosmological constant, albeit its magnitude could be very small. Moreover it is natural to take $m \sim H$ and ${\cal F}$ perhaps even larger, while still below the Planck scale. So both fields $\phi$ and $\hat \phi$ simply reside near their low energy 
minima during inflation. 

Since the inflating universe is almost de Sitter, and the energy scales are high, the tunnelling processes between different vacua of $\phi$ and the discharges of ${\cal Q}$ by membrane emissions will be rapid. These were studied in \cite{KPbubbles} and were found to be very similar to the standard Coleman-De Luccia tunneling processes in the thin-wall approximation \cite{cdl}\footnote{As long as the axion-$4$-form couplings and potentials satisfied certain consistency conditions, which can be readily met.}. It is possible that processes altering the values of $\hat \phi$ and $\hat {\cal Q}$ may also occur, leading to universes with domains of different $M_{Pl}$, which are similar to those noted recently in \cite{bunster}. These processes are however highly suppressed below the Planck scale and so we will ignore them here. Recall that tunneling processes and membrane discharges do not spoil the sequester of QFT vacuum energy contributions. The vacuum energy which is subtracted by the $4$-forms remains safely stored away  from gravity, residing in the non-gravitating $4$-form sector \cite{KPbubbles}. As a result, eternal inflation will generate many regions of the universe with $\phi$ ranging over many low energy minima, with varying
values of the cosmological constant $\la  \frac{M_{Pl}^2 }{m {\cal F} } \, \lambda^4$, just like in 
\cite{Bousso:2000xa}.

As $\Lambda$ drops, the membrane nucleation rate and the $\phi$ tunneling rate will slow down. 
The transitions involving multi-membrane emissions, simultaneously discharging many units of flux,
and jumps in $\phi$ involving large changes in the potential are possible \cite{Brown:2010bc}. 
Hence inflation will not be interrupted adversely by too rapid discharge of vacuum energy.
Nevertheless, inflation should end eventually to allow for reheating, that will avoid the ``empty universe"  problem. 

The ``empty universe'' problem can be avoided in the regions of the inflating universe where 
initially the 
value of $\Lambda$ exceeds that of the inflationary potential \cite{Bousso:2000xa}. Because of this, the light inflaton
is controlled by quantum fluctuations rather than by slow roll. As $\Lambda$ changes due to tunneling and discharges, these patches of the universe will still be in the regime of eternal inflation just before the final jump 
in $\Lambda$, with a random distribution of inflaton values. Right after the last tunneling/discharge event, 
in the interior of the final bubble the cosmological constant will sharply drop to its terminal value, allowing  
slow roll inflation to begin. 
Slow roll inflation will seed the universe with curvature perturbations, and will be followed by the reheating in the geometry with a small final $\Lambda$. 
After the end of inflation, gravity will suppress the transitions to states with large negative cosmological constant \cite{cdl}, implying that the states with small $\Lambda$, positive or negative, will be long lived. In fact sequestering dynamics enhances the longevity of such states \cite{KPbubbles}.

If the scale induced by nonperturbative physics satisfies $\frac{M_{Pl}^2 }{m {\cal F} } \, \lambda^4 \la 10^{-12} \, {\rm eV}^4$ no further discussion would be needed. The superposition of the monodromy and nonperturbative 
gauge dynamics potentials would have many minima, all with vacuum energy bounded by $\frac{M_{Pl}^2 }{m {\cal F} } \, \lambda^4 \la 10^{-12} \, {\rm eV}^4$. Wherever the universe falls, the cosmological constant would be small enough. 

If this is not the case, the value of $\Lambda_{residual}$ can be determined by invoking Weinberg's anthropic argument \cite{Weinberg:1987dv}. Since the step between different $\Lambda_{residual}$ can be arbitrarily small, because for any phase $\theta$ there exist integers $n,l$ such that $\theta +2\pi (l \sigma - n) < \epsilon$, we have that,
up to ${\cal O}(1)$ factors,
\be
\Lambda_{residual} = - \frac{M_{Pl}^2 }{m {\cal F} } \, \lambda^4 \, \sin(\omega l+ \theta)  ~~~~ \to  ~~~~~
\Delta \Lambda_{residual} <  \frac{M_{Pl}^2 }{m {\cal F} } \, \lambda^4 \, \epsilon \, . \label{resccregfinelast}
\ee
Clearly, for integers whose ratios are not good approximants of $\sigma$ the energy differences are much larger. However, assuming the uniform distribution of real numbers, we need $\Delta \Lambda \la 10^{-12} \, {\rm eV}^4$ to ensure that acceptably small values of the observed cosmological constant are not unlikely. Imposing anthropic bounds \cite{Weinberg:1987dv}, the favored value is $\Lambda \simeq 10^{-12} \, {\rm eV}^4$, fitting observations. This value is insensitive to UV physics corrections at 
scales above  $\frac{M_{Pl}^2 }{m {\cal F} } \, \lambda^4$, that remain sequestered away by the topological $4$-form sectors. 

\section{Summary}
\label{sec5}

In \cite{KPlocal,KPbubbles,KPgrav} the author and collaborators have devised a manifestly local theory of vacuum energy sequester, which removes all QFT vacuum energy corrections from the stress energy tensor sourcing gravity. Because the theory is local it admits the standard  Hamiltonian description, and can be used as a starting point for a path integral formulation of gravity coupled to matter. The solutions can have a finite cosmological constant and an infinite worldvolume. The theory involves non-gravitating topological sectors, comprised of $4$-forms whose gauge symmetries allow for Lagrange multiplier fields, that can be used as constraints which ensure diversion of QFT vacuum energy away from $T^{\mu}{}_{\nu}$. The residual cosmological constant however, while radiatively stable, is left completely arbitrary in perturbation theory. 

The dynamics of local vacuum energy sequester has ingredients similar to the decoupling limit of flux monodromy models of large field inflation \cite{KS,NDA}, retaining the nonperturbative dynamics of discharges and tunnelings adjusting fluxes like  \cite{Banks:1991mb}. 
Noting this similarity we have pursued here a `completion' of the local sequester mechanism into a theory with two monodromies.  This route has led to a landscape of the residual cosmological constant, which is desensitized from some of the unknown UV physics. An additional fine structure of vacua, which is scanning the values around zero extremely finely is introduced by coupling one of the fields to a gauge theory, with its period incommensurate with the period of a monodromy.  This reintroduces some radiative instability controlled by the scale of the gauge theory nonperturbative potential, which can be much lower than the cutoff. 

In contrast to \cite{Banks:1991mb}, irrationality of the sequester landscape arises from incommensurability of a monodromy period and a gauge theory period. Since the $4D$ monodromies typically arise in compactifications  \cite{KS,NDA,eva,uranga}, the $4D$ monodromy period and membrane charges are products
of compactified dimensions and so on. This may be able to induce irrational ratios in $4D$. This is similar in spirit to the compactifications considered in \cite{Bousso:2000xa,Feng:2000if,Donoghue:2000fk}. Thus there doesn't seem to be any fundamental obstacle to getting irrational ratios of the periods of $4D$ potentials. Any global symmetries due to them would be emergent, appearing only at low energies. 

Unlike some of the other landscape models, the landscape which we have presented here is purely $4D$. It might arise in special limits 
of compactifications of higher dimensional theories when all internal dimensions are ${\cal O}(1)$ in fundamental units. We do not need a large number of fields or forms, in principle, as long as the ratios of periods are irrational\footnote{It might suffice that the ratios of periods are fractions of large primes, in which case the resulting landscapes span dense discretuums. However such realizations might need a lot of fields and/or forms \cite{Bachlechner:2017zpb}.}.

Resorting to eternal inflation as a colonizing force behind populating the landscape, in a way very similar to 
\cite{Bousso:2000xa}, we see that the late time universes are cosmologies that scan very finely a wide range of values of cosmological constant around zero. Thus one can use Weinberg's anthropic selection to pick the cosmological constant at the observed value of $10^{-12} \, {\rm eV}^4$. Alternatively, if the gauge theory that induces the fine structure of $\Lambda_{residual}$ yields a very suppressed potential term, any vacuum could
immediately have an acceptably small cosmological constant. 

One might object to the pursuits presented here on `philosophical' grounds. We have started with an attempt to
desensitize the cosmological constant from all unknown UV physics, simplistically modeled by loop corrections and dependence on the cutoff. We have ended up with an anthropic landscape. 
Why should we need both of these? Should it not suffice to have only one realized? 

An issue which invariably arises in the decoupling of the unknown UV physics is, how is the renormalized, UV-stabilized cosmological constant value to be determined? The anthropic landscape approach provides a possible answer. The specific realization which we have uncovered shows that the terminal value of the cosmological constant turns out to depend on the protected subsector of the theory -- e.g. the Standard Model -- only very weakly, despite the fact that all local fluctuations with finite wavelengths in this sector obey the standard Equivalence Principle. The terminal cosmological constant is only determined by the dynamics of the sequestering sector, which removes the contributions from the protected subsector form the cosmological constant. This reduces UV sensitivity of $\Lambda_{residual}$ and is a novel and unusual feature by itself. Further, perhaps other ways for choosing the terminal cosmological constant, which do not use the anthropic principle, exist as well. Such frameworks might yield dynamical adjustment of cosmological constant. 

It would be interesting to study the scenario  we have outlined here in more detail, and also pursue closer contact with detailed derivations starting with UV complete frameworks. 
Further,  we recall that the irrational axion proposal of 
\cite{Banks:1991mb} has been primarily designed in the attempt to address the strong CP problem using anthropics. 
It would be interesting to see if such an application can be realized in the context of monodromies and sequester. We cannot refrain from sharing a speculation along these lines, which we find quite curious, in the appendix.

~


{\bf Acknowledgments}: 
NK would like to thank G. D'Amico, A. Lawrence, A. Padilla, J. Terning and  A. Westphal for many 
very useful discussions and collaboration on closely related
ideas, and to A. Hebecker, A. Linde and F. Quevedo for useful comments and questions. NK would like to thank CERN Theory Division for kind hospitality in 
the course of this work. NK is supported in part by the DOE Grant DE-SC0009999.

~


\appendix

\section{Strong CP Problem and Anthropics: a Speculation}

In section (\ref{sec4}) we have assumed that the gauge theory which couples to $\phi$ and generates the fine structure spectrum of the cosmological constant is some gauge theory which is strongly coupled close to 
$M_{GUT}$, with the mass and the monodromy period of $\phi$ being close by. Let us here instead entertain the possibility that the gauge theory is actually just QCD. In this case, $\lambda _{QCD} \simeq 100 \, {\rm MeV}$. This is inspired by \cite{Banks:1991mb}. In a similar vein to \cite{Banks:1991mb}, we keep
$m$ and ${\cal F}$ large. Thus the axion is still very heavy. 

Above the QCD scale there is no QCD-induced correction to $V(\phi)$, which is given only by the monodromy term
$V$. If $\phi$ is heavy, after inflation it will reside in the minima $\phi_{min} = -{\cal Q}/m = - n {\cal F}$. 
Hence the residual cosmological constant, after the vacuum energy contributions are sequestered by $4$-forms, only has to satisfy the constraint that $\Lambda_{residual} < {\rm GeV}^4$ in order not to disturb the normal evolution of the universe down to the QCD confinement scale. 

Supposing this the universe can evolve down to QCD scales, when the QCD instantons generate the additional cosine potential correction\footnote{Again, the cosine is not important; what matters is that the potential is periodic and
differentiable.}  to $V(\phi)$, such that the total potential for $\phi$ becomes 
$V_{tot} = V + \lambda^4_{QCD}[1-\cos(\frac{\omega \phi}{\cal F} + \theta)]$. The phase of the cosine plays the role of the QCD vacuum angle $\theta_{QCD}$. The cosine potential induces a fine structure of the cosmological constant. In particular it shifts slightly 
the minima of $V_{tot}$, solving $\partial_\phi V_{tot} = 0$. This happens during the QCD phase transition, being a part of it since the energy differences between such states are $\la \lambda_{QCD}$. The interim values of $\phi = -n{\cal F}$, set well before the QCD phase transition, will either shift, or tunnel to the new minima. The tunneling rates may be very slow, but since the universe is radiation dominated at that time, the various bubbles relaxing $\phi$ to new nearest minima of $V_{tot}$ can percolate. 

After this the resulting cosmological constant spectrum will be given by Eq. (\ref{resccregfine}), 
with the values of $\phi$ determined by Eq. (\ref{condition}). However in both we now use $\lambda_{QCD}$ instead of $\lambda$. Setting ${\cal U} \sim {\cal U}' \sim \int \hat F/\int F \sim \omega \sim 1$ in the prefactor 
for $\Lambda_{residual}$ yields
$$
\Lambda_{residual} \sim - \frac{M_{Pl}^2 }{m {\cal F} } \, \lambda^4_{QCD} \, \sin(\frac{\omega \phi}{\cal F} + \theta)   \,, ~~~~ - \partial_{\cal Q} V =  \frac{\omega \lambda^4_{QCD}}{m\cal F} \sin(\frac{\omega \phi}{\cal F} + \theta) \, .
$$
Note that $\Lambda_{residual}$ is the total remaining cosmological constant which should now be subjected to anthropic selection to determine its value. So requiring $\Lambda_{residual} \la 10^{-12} \, {\rm eV}^4$ and noting that with our choice of scales $\lambda_{QCD}^4 M_{Pl}^2/m {\cal F} \ga {\rm GeV}^4$, we find that the phase of the sine function in $\Lambda_{residual}$ must be extremely close to an integer multiple of $\pi$:
$$
\frac{M_{Pl}^2 }{m {\cal F} } \, \lambda^4_{QCD} \, |\theta_{QCD} - n\pi| \la 10^{-12} \, {\rm eV}^4 ~~~ \to ~~~ | \theta_{QCD} - n \pi | \la 10^{-44}  \frac{m {\cal F}}{M_{Pl}^2} \, .
$$
Since $\theta_{QCD}$ is defined modulo $2\pi$ this means that $\theta$ is either $0$ or $\pi$ with a very high precision. E.g., if $m$ and ${\cal F}$ are GUT scale, the precision is
$10^{-48}$. If they are lower, the precision is even higher. Putting in a very mild posterior, which is that strong forces do not appear to break CP strongly leads to a prediction: $\theta_{QCD}$ is tiny, being smaller than the current bounds by many orders of magnitude:
$$
\theta_{QCD} < 10^{-44 } \frac{m {\cal F}}{M_{Pl}^2} \, ,
$$
implying that the future searches for the neutron dipole moment will come up empty, if this is the mechanism for
setting $\theta_{QCD} \ll 1$!

Clearly, the precise dependence of $\Lambda_{residual}$ on $\theta_{QCD}$ could be sensitive to other corrections near or above the QCD scale. In the presence of other sectors which confine at higher energies and may also involve CP violations, the bound we found could change. However, the argument presented here leads to so strong a bound on $\theta_{QCD}$ compared to other 
anthropic considerations \cite{Banks:1991mb,anthropictheta} that we could not resist including it.


\begin{thebibliography}{99}

\bibitem{zeldovich} 
Y.~B.~Zeldovich,
JETP Lett.\  {\bf 6}, 316 (1967);
Sov.\ Phys.\ Usp.\  {\bf 11}, 381 (1968).

\bibitem{wilczek} 
F.~Wilczek,
Phys.\ Rept.\  {\bf 104}, 143 (1984).

\bibitem{wein}
S.~Weinberg,
Rev.\ Mod.\ Phys.\  {\bf 61}, 1 (1989).

 
\bibitem{dreitlein} 
J.~Dreitlein,
Phys.\ Rev.\ Lett.\  {\bf 33}, 1243 (1974).
 
\bibitem{linde} 
A.~D.~Linde,
Rept.\ Prog.\ Phys.\  {\bf 42}, 389 (1979).

\bibitem{veltman} 
M.~J.~G.~Veltman,
Phys.\ Rev.\ Lett.\  {\bf 34}, 777 (1975).

\bibitem{dilatons} 
  F.~Coradeschi, P.~Lodone, D.~Pappadopulo, R.~Rattazzi and L.~Vitale,
  JHEP {\bf 1311}, 057 (2013);
  B.~Bellazzini, C.~Csaki, J.~Hubisz, J.~Serra and J.~Terning,
  Eur.\ Phys.\ J.\ C {\bf 74}, 2790 (2014).

\bibitem{witten}
  E.~Witten,
  Int.\ J.\ Mod.\ Phys.\ A {\bf 10}, 1247 (1995);
  Mod.\ Phys.\ Lett.\ A {\bf 10}, 2153 (1995).
  
\bibitem{Hawking:1984hk} 
  S.~W.~Hawking,
  Phys.\ Lett.\  {\bf 134B}, 403 (1984);
  E.~Baum,
  Phys.\ Lett.\  {\bf 133B}, 185 (1983);
  S.~R.~Coleman,
  Nucl.\ Phys.\ B {\bf 307}, 867 (1988);
  Nucl.\ Phys.\ B {\bf 310}, 643 (1988);
  T.~Banks,
  Int.\ J.\ Mod.\ Phys.\ A {\bf 16}, 910 (2001).

\bibitem{Klebanov:1988eh} 
  I.~R.~Klebanov, L.~Susskind and T.~Banks,
  Nucl.\ Phys.\ B {\bf 317}, 665 (1989);
  W.~Fischler, I.~R.~Klebanov, J.~Polchinski and L.~Susskind,
  Nucl.\ Phys.\ B {\bf 327}, 157 (1989).
  
\bibitem{Duff:1989ah} 
  M.~J.~Duff,
  Phys.\ Lett.\ B {\bf 226}, 36 (1989);
  M.~J.~Duncan and L.~G.~Jensen,
  Nucl.\ Phys.\ B {\bf 336}, 100 (1990).

  

\bibitem{Linde:1984ir} 
  A.~D.~Linde,
  Rept.\ Prog.\ Phys.\  {\bf 47}, 925 (1984).
  
\bibitem{Weinberg:1987dv} 
  S.~Weinberg,
  Phys.\ Rev.\ Lett.\  {\bf 59}, 2607 (1987).
 
\bibitem{Vilenkin:1994ua} 
  A.~Vilenkin,
  Phys.\ Rev.\ Lett.\  {\bf 74}, 846 (1995).
  
\bibitem{Banks:1984cw} 
  T.~Banks,
  Nucl.\ Phys.\ B {\bf 249}, 332 (1985).
    
\bibitem{Abbott:1984qf} 
  L.~F.~Abbott,
  Phys.\ Lett.\  {\bf 150B}, 427 (1985).
    
\bibitem{Brown:1987dd} 
  J.~D.~Brown and C.~Teitelboim,
  Phys.\ Lett.\ B {\bf 195}, 177 (1987);
  Nucl.\ Phys.\ B {\bf 297}, 787 (1988).

  
\bibitem{Banks:1991mb} 
  T.~Banks, M.~Dine and N.~Seiberg,
  Phys.\ Lett.\ B {\bf 273}, 105 (1991).
 
         
\bibitem{Bousso:2000xa} 
  R.~Bousso and J.~Polchinski,
  JHEP {\bf 0006}, 006 (2000);
  J.~Polchinski,
  hep-th/0603249.
  
          
\bibitem{Feng:2000if} 
  J.~L.~Feng, J.~March-Russell, S.~Sethi and F.~Wilczek,
  Nucl.\ Phys.\ B {\bf 602}, 307 (2001).

\bibitem{Donoghue:2000fk} 
  J.~F.~Donoghue,
  JHEP {\bf 0008}, 022 (2000);
  Phys.\ Rev.\ D {\bf 69}, 106012 (2004)
  Erratum: [Phys.\ Rev.\ D {\bf 69}, 129901 (2004)].
 
\bibitem{KPglobal}
   N.~Kaloper and A.~Padilla,
  Phys.\ Rev.\ Lett.\  {\bf 112}, no. 9, 091304 (2014);
  Phys.\ Rev.\ D {\bf 90}, no. 8, 084023 (2014)
  [Phys.\ Rev.\ D {\bf 90}, no. 10, 109901 (2014)]; 
  Phys.\ Rev.\ Lett.\  {\bf 114}, no. 10, 101302 (2015).
      
  
\bibitem{KPlocal} 
  N.~Kaloper, A.~Padilla, D.~Stefanyszyn and G.~Zahariade,
  Phys.\ Rev.\ Lett.\  {\bf 116}, no. 5, 051302 (2016).

  
\bibitem{KPbubbles}
  N.~Kaloper, A.~Padilla and D.~Stefanyszyn,
  Phys.\ Rev.\ D {\bf 94}, no. 2, 025022 (2016).
    
\bibitem{KPgrav}
  N.~Kaloper and A.~Padilla,
  Phys.\ Rev.\ Lett.\  {\bf 118}, no. 6, 061303 (2017).
    
\bibitem{etude}
  G.~D'Amico, N.~Kaloper, A.~Padilla, D.~Stefanyszyn, A.~Westphal and G.~Zahariade,
  JHEP {\bf 1709}, 074 (2017).
  

\bibitem{andreimult} 
A.~D.~Linde,
Phys.\ Lett.\ B {\bf 200}, 272 (1988).
  
\bibitem{tseytlin}
A.~A.~Tseytlin,
Phys.\ Rev.\ Lett.\  {\bf 66}, 545 (1991).

\bibitem{HT} 
M.~Henneaux and C.~Teitelboim,
Phys.\ Lett.\ B {\bf 222}, 195 (1989).

\bibitem{Dvali:2001sm} 
  G.~R.~Dvali and A.~Vilenkin,
  Phys.\ Rev.\ D {\bf 64}, 063509 (2001);
  Phys.\ Rev.\ D {\bf 70}, 063501 (2004);
  G.~Dvali,
  Phys.\ Rev.\ D {\bf 74}, 025018 (2006);
  hep-th/0507215.
  
\bibitem{nicolai} 
  A.~Aurilia, H.~Nicolai and P.~K.~Townsend,
  Nucl.\ Phys.\ B {\bf 176}, 509 (1980).
   

  
\bibitem{selft} 
N.~Arkani-Hamed, S.~Dimopoulos, N.~Kaloper and R.~Sundrum,
Phys.\ Lett.\ B {\bf 480}, 193 (2000);
  S.~Kachru, M.~B.~Schulz and E.~Silverstein,
  Phys.\ Rev.\ D {\bf 62}, 045021 (2000).
    
\bibitem{degrav}
  N.~Arkani-Hamed, S.~Dimopoulos, G.~Dvali and G.~Gabadadze,
  hep-th/0209227.
  

   
\bibitem{KS}
  N.~Kaloper and L.~Sorbo,
  Phys.\ Rev.\ Lett.\  {\bf 102}, 121301 (2009);
  Phys.\ Rev.\ D {\bf 79}, 043528 (2009);
  N.~Kaloper, A.~Lawrence and L.~Sorbo,
  JCAP {\bf 1103}, 023 (2011);
  N.~Kaloper and A.~Lawrence,
  Phys.\ Rev.\ D {\bf 95}, no. 6, 063526 (2017).
 
\bibitem{NDA} 
  G.~D'Amico, N.~Kaloper and A.~Lawrence,
  arXiv:1709.07014 [hep-th].
  
\bibitem{andreifernando}  
  J.~J.~Blanco-Pillado, C.~P.~Burgess, J.~M.~Cline, C.~Escoda, M.~Gomez-Reino, R.~Kallosh, A.~D.~Linde and F.~Quevedo,
  JHEP {\bf 0411}, 063 (2004); 
  R.~Kallosh, A.~Linde and B.~Vercnocke,
  Phys.\ Rev.\ D {\bf 90}, no. 4, 041303 (2014).
  
\bibitem{Banks:2010zn} 
  T.~Banks and N.~Seiberg,
  Phys.\ Rev.\ D {\bf 83}, 084019 (2011). 

\bibitem{Bachlechner:2015gwa} 
  T.~C.~Bachlechner,
  Phys.\ Rev.\ D {\bf 93}, no. 2, 023522 (2016). 
  
\bibitem{eva} 
  E.~Silverstein and A.~Westphal,
  Phys.\ Rev.\ D {\bf 78}, 106003 (2008);
  L.~McAllister, E.~Silverstein and A.~Westphal,
  Phys.\ Rev.\ D {\bf 82}, 046003 (2010).

  
\bibitem{uranga} 
  M.~Montero, A.~M.~Uranga and I.~Valenzuela,
  JHEP {\bf 1707}, 123 (2017).

\bibitem{sandora}
  N.~Kaloper and M.~Sandora,
  Phys.\ Rev.\ D {\bf 87}, no. 2, 023531 (2013). 

\bibitem{niven}
I. Niven, {\it Numbers: Rational and Irrational}, Mathematical Association of America: New Mathematical Library (June 1, 1961).

  
\bibitem{milliev}  
  C.~T.~Hill, D.~N.~Schramm and J.~N.~Fry,
  Comments Nucl.\ Part.\ Phys.\  {\bf 19}, no. 1, 25 (1989);
  J.~A.~Frieman, C.~T.~Hill and R.~Watkins,
  Phys.\ Rev.\ D {\bf 46}, 1226 (1992);
  M.~Fukugita and T.~Yanagida,
  YITP-K-1098;
  J.~A.~Frieman, C.~T.~Hill, A.~Stebbins and I.~Waga,
  Phys.\ Rev.\ Lett.\  {\bf 75}, 2077 (1995)
  Y.~Nomura, T.~Watari and T.~Yanagida,
  Phys.\ Rev.\ D {\bf 61}, 105007 (2000).
   
   
\bibitem{vafawitten} 
  C.~Vafa and E.~Witten,
 {Phys.\ Rev.\ Lett.\  {\bf 53}, 535 (1984)}.
  
\bibitem{exercises}
  X.~Dong, B.~Horn, E.~Silverstein and A.~Westphal,
  Phys.\ Rev.\ D {\bf 84}, 026011 (2011). 


\bibitem{holdom} 
  B.~Holdom,
  Phys.\ Lett.\  {\bf 166B}, 196 (1986).
  
\bibitem{cdl} 
  S.~R.~Coleman and F.~De Luccia,
  Phys.\ Rev.\ D {\bf 21}, 3305 (1980).

\bibitem{bunster} 
  C.~Bunster and A.~Gomberoff,
  Phys.\ Rev.\ D {\bf 96}, no. 2, 025013 (2017).
  
\bibitem{Brown:2010bc} 
  A.~R.~Brown and A.~Dahlen,
  Phys.\ Rev.\ D {\bf 82}, 083519 (2010);
  M.~Kleban, K.~Krishnaiyengar and M.~Porrati,
  JHEP {\bf 1111}, 096 (2011).
  
\bibitem{guido} 
  G.~D'Amico, R.~Gobbetti, M.~Schillo and M.~Kleban,
  Phys.\ Lett.\ B {\bf 725}, 218 (2013);
  JCAP {\bf 1303}, 004 (2013).
  
  

  
\bibitem{Bachlechner:2017zpb} 
  T.~Higaki and F.~Takahashi,
  Phys.\ Lett.\ B {\bf 744}, 153 (2015);
 T.~C.~Bachlechner, K.~Eckerle, O.~Janssen and M.~Kleban,
  arXiv:1703.00453 [hep-th];
  JHEP {\bf 1711}, 036 (2017). 


\bibitem{anthropictheta} 
  N.~Weiss,
  Phys.\ Rev.\ D {\bf 37}, 3760 (1988);
  F.~Takahashi,
  Prog.\ Theor.\ Phys.\  {\bf 121}, 711 (2009);
  N.~Kaloper and J.~Terning,
  arXiv:1710.01740 [hep-th];
  M.~Dine, L.~Stephenson Haskins, L.~Ubaldi and D.~Xu,
  arXiv:1801.03466 [hep-th].

           
\end{thebibliography}
\end{document}